\documentclass[preprint]{aastex}

\newcommand{\asca}{{\sl ASCA}}
\newcommand{\rosat}{{\sl ROSAT}}
\newcommand{\rxte}{{\sl RXTE}}
\newcommand{\snr}{SN~1006}
\newcommand{\vhe}{very high energy}

\slugcomment{Accepted for publication in ApJ.}

\shorttitle{X-Ray Synchrotron Emission from \snr}
\shortauthors{Allen et~al.}

\begin{document}

\title{X-Ray Synchrotron Emission from 10--100 TeV Cosmic-Ray Electrons \\
   in the Supernova Remnant \snr}


\author{G.\ E.\ Allen}
\affil{MIT Center for Space Research, 77 Massachusetts Avenue, NE80-6029, 
   Cambridge, MA 02139-4307}

\author{R.\ Petre}
\affil{NASA/Goddard Space Flight Center, Laboratory for High Energy
   Astrophysics, Code 662, Greenbelt, MD 20771}

\and

\author{E.\ V.\ Gotthelf}
\affil{Columbia Astrophysics Laboratory, Columbia University, Pupin Hall, 
   550 West 120th Street, New York, NY 10027}


\begin{abstract}

We present the results of a joint spectral analysis of \rxte\ PCA, \asca\ SIS, 
and \rosat\ PSPC data of the supernova remnant \snr.  This work represents 
the first attempt to model both the thermal and nonthermal X-ray emission 
over the entire X-ray energy band.  The thermal flux is described by a 
nonequilibrium ionization model with an electron temperature $kT_{e} = 
0.6$~keV, an ionization timescale $n_{\rm 0}t = 9 \times 10^{9}$~cm$^{-3}$~s, 
and a relative elemental abundance of silicon that is 10--18 times larger 
than the solar abundance.  The nonthermal X-ray spectrum is described by a 
broken power law model with low- and high-energy photon indices $\Gamma_{1} 
= 2.1$ and $\Gamma_{2} = 3.0$, respectively.  Since the nonthermal X-ray 
spectrum steepens with increasing energy, the results of the present 
analysis corroborate previous claims that the nonthermal X-ray emission is 
produced by synchrotron radiation.  We argue that the magnetic field 
strength is significantly larger than previous estimates of about 10~$\mu$G 
and arbitrarily use a value of 40~$\mu$G to estimate the parameters of the 
cosmic-ray electron, proton, and helium spectra of the remnant.  The results 
for the ratio of the number densities of protons and electrons 
($R = 160$ at 1~GeV), the 
total energy in cosmic rays ($E_{\rm cr} = 1 \times 10^{50}$~ergs), and the 
spectral index of the electrons at 1~GeV ($\Gamma_{e} = 2.14 \pm 0.12$) 
are consistent with the hypothesis that Galactic cosmic rays are accelerated 
predominantly in the shocks of supernova remnants.  Yet, the remnant may or 
may not accelerate nuclei to energies as high as the energy of the ``knee,'' 
depending on the reason why the maximum energy of the electrons is only 
10~TeV.

\end{abstract}



\section{Introduction}
\label{int}

The search for evidence of the origin of Galactic cosmic rays has been an 
active area of research for many decades.  While little evidence exists 
about the sites at which \vhe\ nuclei are accelerated, the results of recent 
X-ray and gamma-ray observations indicate that at least some of the 
cosmic-ray electrons are accelerated in the shocks of supernova remnants 
\citep{koy95,koy97,all97,tan98,all99,vin99,sla99,mur00,bor00,dye01}.  For 
example, \snr\ is one remnant for which there is evidence that cosmic-ray 
electrons have been accelerated to energies as high as about 100~TeV 
\citep{koy95,tan98}.  In this paper, measurements of the X-ray and 
radio emission of the remnant are used to determine the parameters of the 
nonthermal electron spectrum.  Although there is no evidence to indicate 
that cosmic-ray nuclei are accelerated in \snr, relativistic electrons and 
nuclei are expected to be accelerated in a similar manner \citep{ell91}.  
Therefore, we estimate the parameters of the proton and helium spectra of 
the remnant.  The results of this analysis show that the remnant is a 
significant source of Galactic cosmic rays (at least cosmic-ray electrons) 
and provide some support for the idea that Galactic cosmic rays are 
accelerated predominantly in the shocks of supernova remnants.

Section~\ref{obs} contains a description of some of the results of previous 
analyses of the X-ray data of \snr.  The data and analysis techniques used 
for the present work are described in section~\ref{dat}.  Section~\ref{dis} 
includes a discussion of the results of the present analysis.  The 
conclusions are reviewed in section~\ref{con}.


\section{Observational X-Ray Record}
\label{obs}

The first detection of X-ray emission from \snr\ was reported by \citet{win76}.
The results of analyses of X-ray spectral data of the remnant have been 
reported by many authors \citep{win76,zar79,win79,too80,bec80,pye81,gal82,
var85,ham86,koy87,koy95,lea91,oza94,wil96,rey96,lam98,vin00,dye01}.  As 
early as 1979, it was realized that the global X-ray spectrum of \snr\ cannot 
be fitted with a single thermal or nonthermal model 
\citep{win79}.  This realization is not surprising in retrospect because 
observations of the remnant with imaging X-ray devices show that the X-ray 
spectrum varies as a function of position in the remnant.  The high-energy 
X-ray image of the remnant, like the radio image, is dominated by emission 
from the northeastern and southwestern rims \citep{pye81,wil96,win97}.  Since 
the spectra of the rims are more or less featureless, 
the emission from these regions is reported to be nonthermal \citep{koy95,
rey96,dye01}.  At energies of less than 0.5~keV, a significant 
amount of emission is observed from the interior of the remnant, especially 
in the southwest \citep{pye81,wil96}.  Since the X-ray spectrum of 
the interior exhibits atomic emission-line features, a substantial amount of 
the emission from this region must be thermal bremsstrahlung.

Several of the spectral analyses of \snr\ included two spectral components.  
Invariably, one of the spectral components is thermal emission from an 
optically thin plasma.  Such a model is needed to describe the emission-line 
features of oxygen, neon, magnesium, silicon, and perhaps iron 
\citep{koy95,vin00,dye01}.  Some analyses have included an additional 
thermal component, while others have included a nonthermal component.  For 
example, Hamilton et~al.\ (1986) modeled the spectrum using two thermal 
components.  One component is used to describe the forward-shocked 
interstellar medium, and the other describes the reverse-shocked ejecta.  To 
explain the nearly featureless nature of the global X-ray spectrum of the 
remnant, they used a model for the shocked ejecta in which most of the 
emission is produced by a nearly pure layer of fully ionized carbon.  The 
elements more massive than carbon are assumed to be significantly underionized 
in the ejecta, to suppress the X-ray line emission that would 
otherwise be produced by these elements.  However, these assumptions are 
inconsistent with the results of subsequent analyses.  The results of an 
analysis of the ultraviolet absorption-line features along the line of sight 
to the S-M star suggest that most of the ejected silicon has already been 
shocked \citep{ham97}, and therefore may not be in a low ionization state.  
This result is supported by X-ray observations that reveal line emission 
from several highly ionized elements: oxygen, neon, magnesium, silicon, and
perhaps iron \citep{koy95,vin00,dye01}.  Since the 
relative abundance of the shocked silicon producing the X-ray emission is 
much larger than the relative abundance of silicon in the solar system, most 
of the shocked silicon is ejecta.  Furthermore, the results of the analysis 
of \cite{lam98} suggest that the total amount of reverse-shocked carbon 
required by the model of Hamilton et~al.\ may be 
unreasonably large for the remnant of a type Ia supernova (i.e., may be well 
in excess of the Chandrasekhar mass).  For these reasons, the thermal model 
of Hamilton et~al.\ provides a poor description of the 
X-ray emission of \snr.

\citet{wil96} present a nonthermal interpretation of the high-energy X-ray 
emission.  They argue that the X-ray emission from the rims is produced by 
relativistic electrons that are beamed from an unobserved central compact 
object.  This conclusion seems to be based largely on the X-ray morphology 
of \snr, because no evidence of a compact object or of particle jets in the 
remnant has been reported.  For comparison, the pulsar and the pulsar jet of 
the Vela supernova remnant are observed to produce radio, X-ray, and 
gamma-ray emission \citep{mar95,yos97,boc98}.  
\cite{wil96} set an upper limit on the flux of a point source of $6.3 \times 
10^{-14}$~erg~cm$^{-2}$~s$^{-1}$.  The corresponding constraint on the 
blackbody temperature of a 10~km neutron star is $T_{\rm BB} < 6.9 \times 
10^{5}$~K (at a distance of 1.7~kpc).  Some of the models of cooling neutron 
stars indicate that the temperature of a neutron star is higher than this 
upper limit at an age of 990~yr
\citep{sla95}.  Furthermore, the sketchy record of the optical light curve 
of the supernova \citep{sch96}, the lack of evidence of an OB association 
near \snr\ (Hamilton et~al.\ 1986), the large distance of the remnant from 
the plane 
of the Galaxy [$z = +430 (d / 1.7\ {\rm kpc})$~pc], measurements of the 
average expansion rate of the remnant \citep{mof93}, and measurements of 
the relative abundances of the elements producing the ultraviolet absorption 
lines \citep{fes88} suggest that \snr\ was produced by a type Ia supernova.  
Type Ia supernovae are generally not expected to produce compact objects.  
For these reasons, it seems unlikely that \snr\ contains an 
undetected compact object that is responsible for the high-energy X-ray 
emission from the rims.

\citet{koy95} advance an alternative nonthermal explanation of the 
high-energy X-ray emission.  They conclude that the emission from the rims 
is synchrotron radiation from electrons that have been accelerated to 
energies of about 100~TeV.  \cite{rey96} argues convincingly that this 
conclusion is the only plausible one and shows that the nonthermal X-ray 
emission of the rims can be described by a model in which both the radio and 
X-ray emission are produced by a common population of shock-accelerated 
electrons.  The strong correlation of the features of a high-energy X-ray 
image and a radio image of the northeastern rim of \snr\ \citep{win97} 
supports this result.  Furthermore, the detection of inverse-Compton TeV 
gamma-ray emission from the northeastern rim of the remnant confirms the 
presence of \vhe\ electrons in \snr\ \citep{tan98}.  Since the TeV 
photons are produced by electrons that have energies similar to the energies 
needed to produce X-ray synchrotron radiation and since the radio, X-ray, 
and gamma-ray spectral data can be fitted with simple models of the spectrum of 
shock-accelerated electrons \citep{mas96,aha99}, the conclusion 
that the nonthermal X-ray emission is produced by synchrotron radiation 
seems unavoidable.  


\section{Data and Analysis}
\label{dat}

To accurately model the shape of the entire X-ray spectrum of \snr, from 
0.12 to 17~keV, we analyzed data obtained using the Proportional Counter Array 
(PCA) of the {\sl Rossi X-Ray Timing Explorer} (\rxte) satellite, the 
Solid-state Imaging Spectrometers (SISs) of the {\sl Advanced Satellite for 
Cosmology and Astrophysics} (\asca), and  the Position Sensitive 
Proportional Counter (PSPCB, hereafter just PSPC) of the {\sl R\"{o}ntgen} 
satellite (\rosat).  Between 1996 February 18 and 1996 February 20, \snr\ 
was observed for 21~ks at a location on the northeastern rim ($\alpha_{2000} 
= 15^{\rm h} 4\fm0$, $\delta_{2000} = -41\arcdeg 48\arcmin$) and for 20~ks 
at a location on the southwestern rim ($\alpha_{2000} = 15^{\rm h} 1\fm8$, 
$\delta_{2000} = -42\arcdeg 6\arcmin$) using the PCA.  The PCA is a 
spectrophotometer that is comprised of an array of five coaligned proportional 
counter units that are mechanically collimated to have a field of view of 
$1\arcdeg$~FWHM \citep{jah96}.  The array is sensitive to photons that have 
energies between about 2 and 60~keV and has an energy resolution $\Delta E / 
E = 0.18$ at 6~keV.  The maximum on-axis collecting area is about 
7000~cm$^{2}$.  The PCA data were screened to remove the time intervals 
during which (1) one or more of the five proportional counter units is off, 
(2) \snr\ is less than $10\arcdeg$ above the limb of the Earth, (3) the 
background model is not well defined, and (4) the pointing direction of the 
detectors is more than
$0\fdg02$ from the nominal pointing direction in either right ascension or 
declination.  The set of X-ray events that satisfies these four selection 
criteria includes 7~ks of the data of the northeastern pointing and 11~ks 
of the data of the southwestern pointing.  Spectra were constructed for 
each of the two PCA pointings.  The two spectra are quite similar because 
the field of view of the PCA is large enough to include the entire remnant 
at both pointing locations.  Therefore, the PCA data are insensitive to 
modest changes in the shape of the spectrum of the remnant as a function of 
position in the remnant.  The PCA background spectrum for \snr\ was 
estimated using version 1.5 of the FTOOL {\sl pcabackest}\footnote{For more 
information about FTOOLS, see 
http://heasarc.gsfc.nasa.gov/docs/software/ftools/ftools\_menu.html.}.  
This version of {\sl pcabackest} includes an estimate of the charged-particle 
and diffuse cosmic X-ray backgrounds based on the ``VLE'' count rate during 
observations of ``source-free'' regions of the sky and an estimate of the 
background associated with the decay of radioactive material that is 
activated when the spacecraft passes through the South Atlantic Anomaly.  
The tool has been found to produce slightly inaccurate estimates of the 
background associated with astrophysical sources.  For example, when the 
background spectrum is subtracted from the spectral data of \snr, the 
resulting ``source'' spectrum of the remnant has a negative count rate at 
high energies.  To compensate for this problem, the spectrum of the 
background associated with the South Atlantic Anomaly is multiplied by a 
factor of 0.2.  This factor is determined by requiring the average 
difference between the source spectrum and the source model to be zero at 
energies between 27 and 71~keV.  In this energy band, the total number of 
events associated with \snr\ is expected to be negligibly small compared to 
the number of events associated with the background.  The adjustment reduces 
the count rate of the 2--10~keV activation background from 1.0
to 0.2~counts~s$^{-1}$ for the top layer of anodes in the PCA.  After 
applying this adjustment, the average 2--10~keV background count-rate for 
the top layer is estimated to be $17.2 \pm 0.1 $~counts~s$^{-1}$ for both 
pointings.  

On 1993 August 19-20, 1993 September 13-15, and 1996 February 20-21, the 
instruments on \asca\ \citep{tan94} were used to observe \snr\ for 43~ks at 
a location in the center of the remnant ($\alpha_{2000} = 15^{\rm h} 2^{\rm m} 
48^{\rm s}$, $\delta_{2000} = -41\arcdeg 55\arcmin 46\arcsec$), for 38~ks at 
a location on the northeastern rim ($\alpha_{2000} = 15^{\rm h} 3^{\rm m} 
32^{\rm s}$, $\delta_{2000} = -41\arcdeg 46\arcmin 25\arcsec$), and for 57~ks 
at a location on the southwestern rim ($\alpha_{2000} = 15^{\rm h} 2^{\rm m} 
34^{\rm s}$, $\delta_{2000} = -42\arcdeg 2\arcmin 58\arcsec$), respectively.  
The SIS detectors are CCD devices \citep{bur94} sensitive to photons that 
have energies between about 0.5 and 10~keV.  The devices have an energy 
resolution $\Delta E / E = 0.06$--0.08 (FWHM) at 1.5~keV and a 
field of view of $22' \times 22'$.  The spatial resolution of the telescope 
$\lesssim 1.5'$ (50\% encircled-photon radius) for the SIS data used here.  
The maximum on-axis effective area of the system is about 220~cm$^{2}$.  The 
SIS data were screened using the standard ``REV2'' criteria that exclude time 
intervals (1) associated with passages through the South Atlantic Anomaly 
and regions of relatively high particles fluxes, (2) during which the bright 
limb of the Earth is in the field of view, and (3) during which the source 
is occulted by the Earth.  The set of X-ray events that satisfies these 
selection criteria includes 23, 20, and 43~ks of data for the center, 
northeastern rim, and southwestern rim, respectively.  The SIS spectra for 
these three spatially separate regions, which correspond to the central,
northeastern, and southwestern ovals of Figure~\ref{fig1}, have event rates 
of 0.21, 0.89, and 0.66~counts~s$^{-1}$, respectively.  

On 1993 January 28--29, \snr\ was observed for 6~ks at a location near 
the center of the remnant ($\alpha_{2000} = 15^{\rm h} 2^{\rm m} 55^{\rm s}$, 
$\delta_{2000} = -41\arcdeg 55\arcmin 12\arcsec$) using the PSPC.  The PSPC 
is a multiple-wire proportional counter sensitive to photons that have 
energies between about 0.1 and 2~keV.  The energy resolution $\Delta E / E = 
0.43$ at 0.93~keV.  The telescope has a field of view of $2^{\arcdeg}$.  The
spatial resolution $\lesssim 30\arcsec$ (50\% encircled-photon radius) for
the PSPC data used here.  The maximum on-axis effective area of the system is 
about 260~cm$^{2}$ \citep{pfe87}.  Spectra were constructed using the PSPC 
data for three spatially separate regions: the bright northeastern rim, the 
bright southwestern rim, and the center of the remnant (the three 
rectangular regions of Fig.~\ref{fig1}).  The total PSPC event rates for 
these three spectra are 5.6, 6.1, and 11.9~counts~s$^{-1}$, respectively.  
The PSPC background for \snr\ is negligibly small.  \cite{plu93} report that 
the event rate associated with unrejected cosmic-ray particles is $4 \times 
10^{-6}$~counts~s$^{-1}$~keV$^{-1}$~arcmin$^{-2}$ in the pulse-height range 
$18 \le {\rm PHA} \le 249$.  In this case, the cosmic-ray event rate over 
the entire remnant is only about $3 \times 
10^{-3}$~counts~s$^{-1}$~keV$^{-1}$.  Since this rate is considerably 
smaller than the PSPC count-rate for \snr\ at all energies from 0.12 to 2~keV, 
the backround is neglected.  

A joint fit to the \rxte\ PCA, \asca\ SIS, and \rosat\ PSPC data was performed
using version 10.0 of the spectral-fitting software package XSPEC\footnote{For 
more information about XSPEC, see
http://heasarc.gsfc.nasa.gov/docs/xanadu/xspec/index.html.}.  Since 
\citet{koy95} and others report evidence of both thermal and nonthermal 
emission, the spectral fits were performed using several combinations of 
thermal and nonthermal emission components.  The thermal components include 
a bremsstrahlung model (Brem), a model of a thin thermal plasma in 
collisional ionization equilibrium \citep[RS;][]{ray77}, and a model of a thin 
thermal plasma that is not in ionization equilibrium \cite[NEI;][]{ham83}.  
The nonthermal components include power law (PL) and broken power law (BPL) 
models.  For each fit, only the absolute normalization of each spectral 
component was allowed to vary from one region to another.  While the spectra 
most likely vary as a function of position in the remnant, we find no 
evidence of a significant variation in the shape of the nonthermal spectra 
of the different regions.  Furthermore, \citet{koy95} and \citet{vin00} report 
no evidence of spatial variation in the emission of the interior of the 
remnant.  

Table~\ref{tab1} summarizes the results of some of the fits.  Note that the 
values of $\chi^{2}/\nu$ $> 1$ for all of the fits.  This result suggests 
that the value of $\chi^{2}$ may be sensitive to systematic errors in the 
responses of the detectors as well as differences between the spectral 
models and the spectral data.  Therefore, the $\chi^{2}$ test statistic does 
not provide an unambiguous measure of the goodness of fit of the models.  
However, if the contributions of the systematic errors to $\chi^{2}$ do not 
change from one fit to another, the differences in $\chi^{2}$ provide a 
useful measure of the relative goodness of fit of the different models.  In 
this case, $\Delta \chi^{2} = \chi^{2} - \chi^{2}_{\rm min}$ is distributed 
as a $\chi^{2}$ variable with the number of degrees of freedom equal to the 
number of free parameters in the fit \citep{lam76}.  For example, 26
free parameters are used for the best-fit model of Table~\ref{tab1}.  
Therefore, the values of $\Delta \chi^{2} = 28.9$, 35.6, 39.3, and 50.6 are 
expected to correspond to the 1~$\sigma$, 90\%, 2~$\sigma$, and 3~$\sigma$ 
confidence levels, respectively.  These values of $\Delta \chi^{2}$ are used 
in this paper to determine the confidence intervals of the fit and inferred 
parameters.  Since the differences between the value of $\chi^{2}$ for the 
BPL$+$NEI model and the values of $\chi^{2}$ for the other sets of models 
$\ge 93$, the BPL$+$NEI model seems to provide a significantly better fit to 
the entire broadband spectral data than the other sets of models listed in 
Table~\ref{tab1}.  Table~\ref{tab2} lists the fit parameters of the BPL$+$NEI 
model.  The spectral data and the best-fit model are shown in 
Figures~\ref{fig2}--\ref{fig4}.  In general, the model fits the 0.12--17~keV 
data quite well.  

The thermal component of the best-fit model is a nonequilibrium ionization 
model of a thin thermal plasma that has different electron and ion 
temperatures (Hamilton et~al.\ 1983).  The shape of the X-ray continuum is 
specified 
by the temperature associated with the forward shock, $T_{\rm fs}$, and the 
parameter $n_{0}^{2}E_{0}$, where $n_{0}$ is the ambient density of 
interstellar hydrogen and $E_{0}$ is the initial kinetic energy of the ejecta.
This model is implemented in XSPEC version 10.0 using a fixed grid in the 
$T_{\rm 
fs}$-$n_{0}^{2}E_{0}$ plane.  The temperature is quantized by factors of 
$10^{0.25}$ from $10^{6.25}$ to $10^{8.25}$~K and $n_{0}^{2}E_{0}$ is 
quantized by factors of 10 from $10^{48}$ to $10^{53}$~ergs~cm$^{-6}$.  The 
best-fit values of these two parameters are $T_{\rm fs} = 10^{8.0}$~K and 
$n_{0}^{2}E_{0} = 10^{50}$~ergs~cm$^{-6}$ (Fig.~\ref{fig5}).  Figure~\ref{fig5} 
includes curves along which the ionization timescale, $n_{0}t$, and the 
characteristic electron temperature, $kT_{e}$, are constant.  The 
elongated confidence level contours in this figure nearly lie along curves 
of constant ionization timescale.  The best-fit values of $T_{\rm fs}$ and 
$n_{0}^{2}E_{0}$ are driven by the thermal emission of \snr\
because the same values for these two parameters (within the statistical 
uncertainties) are obtained if the same set of spectral models is fitted to 
only the \asca\ SIS spectrum of the center of the remnant.  The relative 
elemental abundances of the elements oxygen, neon, magnesium, and silicon 
are included as free parameters in the fit because the \asca\ SIS spectrum 
of the central region of \snr\ exhibits atomic emission-line features 
associated with these elements \citep{koy95}.  The abundances of sulphur 
and calcium are fixed to be the same as the abundance of silicon because 
this situation is more or less consistent with the type Ia supernova model 
of \citet{nom84}.  Since the value of $\chi^{2}$ is not significantly reduced 
if the abundances of helium, carbon, nitrogen, iron, or nickel are included 
as free parameters in the fit, these five abundances are set to be the same 
as the relative abundances of these elements in the  solar system 
\citep{and89}.  The fitted abundances of oxygen and magnesium are 
consistent with the solar abundances of these elements.  The abundance of 
neon in \snr\ is less than 80\% of the relative solar abundance of neon.  The 
fitted relative abundance of silicon (and sulphur and calcium) is 10--18 
times larger than the solar abundance of silicon. The uncertainties of the 
abundances are somewhat larger than the statistical uncertainties specified 
in Table~\ref{tab2} because (Hamilton et~al.\ 1983) note that the 
intensities used in the NEI model for the strong emission lines are 
generally uncertain by at least a factor of 2.

A broken power law model is used here to approximate a gradually steepening 
nonthermal X-ray spectrum.  Such a spectrum is expected if the nonthermal 
emission is produced by synchrotron radiation \citep{rey96}.  As shown in 
Figure~\ref{fig6}, the nonthermal spectrum of \snr\ steepens significantly 
with increasing energy.  The difference between the high-energy and low-energy 
photon indices of the broken power law $\ge 0.7$ at the 3~$\sigma$ 
confidence level.  The value of the break energy of the broken power law 
($E_{b} = 1.85 \pm 0.18$~keV) is not physically meaningful.  It does not 
correspond to a feature in the nonthermal X-ray spectrum of \snr.  This 
value is determined by the transition from the relatively high count rate 
portion of the \rosat\ PSPC spectrum to the relatively high count rate 
portion of the \rxte\ PCA spectrum.


\section{Discussion}
\label{dis}

\subsection{Thermal Emission}

Only one thermal emission model is used in the present analysis to describe 
the properties of the X-ray--emitting material in \snr.  The use of one 
thermal component is undoubtedly an oversimplification, because both the 
forward-shocked interstellar material and the reverse-shocked ejecta produce 
X rays.  The large fitted abundance of silicon suggests that most of the 
shocked silicon is ejecta.  Yet the inferred amount of swept-up mass implies 
that most of the shocked mass is interstellar.  For these reasons, the use 
of a single thermal model to describe the global thermal emission properties 
of the remnant may provide a poor description of any given volume element of 
the X-ray--emitting material.  Nevertheless, the use of a single thermal 
model is sufficient to accurately characterize the properties of the 
nonthermal X-ray emission (our primary goal), because the X-ray emission of 
the remnant is dominated by nonthermal emission at energies above 1~keV,
because the results for the broken power law component are 
insensitive to the details of the single thermal component used in the fits 
(compare the results obtained for the broken power laws of the ``BPL$+$NEI'' 
and ``BPL$+$RS'' models listed in Table 1), and because \cite{koy95} and 
\cite{vin00} report no evidence of spatial variation of the thermal X-ray 
spectrum of the interior of \snr.

Models that include two thermal components and one nonthermal component 
were fitted to the data.  However, the results for the fit parameters of these 
three-component models showed that the parameters of one of the two thermal 
components are essentially unconstrained.  

\cite{vin00} favor a model for the X-ray emission of the interior of \snr\ 
that includes two thermal components instead of one thermal component and 
one nonthermal component.  They find that the temperatures of the hotter 
thermal components $kT_{e} > 3$~keV for both the northern and southern 
regions of the interior.  These temperatures are quite high for the observed 
Galactic supernova remnants.  Of the young, shell-type remnants, only 
Cassiopeia~A is reported to have a thermal plasma with a temperature this 
high.  \citet{ham84} show that it may be possible for a reverse-shocked 
silicon-dominated plasma to have electron temperatures this high, but such a 
metal-rich plasma would cool quickly.  Therefore, the cooling time would 
have to be comparable to the age of the remnant or longer.  Otherwise, the 
reverse-shocked electrons may be too cool to produce much X-ray emission.  
The model of \cite{vin00} cannot explain nonthermal X-ray emission from 
the interior of the remnant.  Since \cite{dye01b} show that about 25\% of 
the radio emission at 843~MHz is produced in the interior of \snr, a similar 
fraction of nonthermal X-ray synchrotron emission is expected from this 
region.  Our fits to the X-ray data with a model that includes one thermal 
and one nonthermal component suggest that $13 \pm 7$\% of the nonthermal 
X-ray emission is produced in the region.  Since this fraction is comparable 
to the fraction of the radio emission from the interior of \snr, we favor a 
model for the interior (and the rims) that includes one 
thermal component and one nonthermal component instead of two thermal 
components.  It should be emphasized that the principal results of the 
present analysis concerning the properties of the cosmic rays in \snr\ are 
insensitive to this choice, since only a small fraction of the high-energy 
X-ray emission is produced in the interior of the remnant.  

The parameters of the best-fit model are listed in Table~\ref{tab2}.  
Table~\ref{tab3} includes a list of the inferred values of the 
characteristic energy of the thermal electrons, $kT_{e}$ (see 
Fig.~\ref{fig5}), the velocity of the forward shock, $v_{\rm fs}$, the 
radius of the forward shock, $r_{\rm fs}$, the distance of the remnant, $d$, 
the ionization timescale, $n_{0}t$ (see Fig.~\ref{fig5}), the age of the 
remnant, $t$, the density of interstellar hydrogen, $n_{0}$, and the swept-up 
mass, $M_{s}$ (Hamilton et~al.\ 1983).  These inferences are based on the 
assumptions that \snr\ is in a Sedov phase, that the initial kinetic energy 
of the ejecta is $10^{51}$~ergs, that the X-ray emission is dominated by 
forward-shocked interstellar material, and that the distribution of the 
ambient material around \snr\ is homogeneous.

The best-fit value for the absorption column density of interstellar hydrogen 
$n_{\rm H} = (5.6 \pm 0.6) \times 10^{20}$~atoms~cm$^{-2}$ (at the 90\% 
confidence level).  This value is consistent with the result of Schaefer 
(1996) [$n_{\rm H} = (6.1 \pm 0.6) \times 10^{20}$~atoms~cm$^{-2}$], who 
analyzed a compilation of results from several sources.  The relative 
elemental abundances of oxygen, neon, magnesium, and silicon are consistent 
with the results of \cite{koy95}, but Vink et~al.\ (2000) obtain a somewhat 
lower abundance for oxygen.  The inferred energy of the thermal electrons 
($kT_{e} = 0.58^{+0.02}_{-0.27}$~keV at the 90\% confidence level) is 
similar to the results obtained by Vink et~al.\ (2000; $kT_{e} = 0.71 
\pm 0.15$ and $0.78 \pm 0.09$~keV) and Dyer et~al.\ (2001a;
$kT_{e} = 0.60^{+0.70}_{-0.52}$), but it is somewhat lower than the 
result of Koyama et~al.\ (1995; $kT_{e} = 1.6$~keV).  The results of an 
analysis of the ultraviolet emission-line data of the northwestern rim 
favor a low electron temperature ($kT_{e} < 0.05 kT_{p} = 0.05 
\case{3}{16} m_{p} v_{\rm fs}^{2} = 0.7$~keV), but the results cannot 
exclude temperatures as high as $kT_{e} = 2.9$~keV \citep{lam96}.  The 
inferred shock velocity ($v_{\rm fs} = 2700^{+200}_{-1700}$~km~s$^{-1}$ at 
the 90\% confidence level) is consistent with the results of Kirshner, 
Winkler, \& Chevalier (1987; $v_{\rm fs} = 2800$--3870~km~s$^{-1}$), Raymond, 
Blair, \& Long (1995; $v_{\rm fs} = 2300$~km~s$^{-1}$), and Laming et~al.\ 
(1996; $v_{\rm fs} = 2600 \pm 300$~km~s$^{-1}$) for the northwestern rim of 
\snr, but \cite{dwa98} argue that the average forward shock velocity is 
$4000 \pm 500$~ks~s$^{-1}$.  

The inferred distance ($d = 1.4^{+2.3}_{-0.1}$~kpc at the 90\% confidence 
level) is consistent with the result of Schaefer (1996; $d = 1.59 \pm 
0.13$~kpc) and with a distance estimate that is based on the velocity and 
proper motion of the forward shock in the northwest.  As described above, 
\cite{lam96} estimate that the velocity of the forward shock in the northwest 
$v_{\rm fs} = 2600 \pm 300$~km~s$^{-1}$.  Since neutral atoms have very 
short lifetimes after having passed through the shock into the hot 
postshock gas, the H$\alpha$ filaments trace the location of the forward 
shock \citep{che78,lon88,win97}.  The results of 
analyses of the mean proper motion of the filaments along the northwestern 
rim of \snr\ suggest that $\bar{\mu} = 0\farcs30 \pm 0\farcs04$~yr$^{-1}$ 
(Long et~al.\ 1988) or $\bar {\mu} = 0\farcs39 \pm 0\farcs06$~yr$^{-1}$ 
\citep{hes81}.  The weighted average of these two results is $0\farcs33 \pm 
0\farcs03$~yr$^{-1}$.  Therefore the implied distance of \snr\ $d = v_{\rm fs} 
/ \langle \bar{\mu} \rangle = 1.66 \pm 0.24$~kpc.  These distance estimates 
are consistent with the upper and lower limits on the distance \citep[Hamilton
et~al.\ 1986, 1997;][]{wu93,bur00}.  Collectively, the results suggest 
that $1.4 \le d \le 2.0$~kpc.  

The ionization timescale ($n_{0}t = 240$--650~cm$^{-3}$~yr), age ($t = 
760$--6300~yr), and ambient density of interstellar hydrogen ($n_{0} = 
0.10$--0.37~cm$^{-3}$), are consistent with the ionization timescale of 
Koyama et~al.\ (1995; $n_{0}t = 300$~cm$^{-3}$~yr), the known age of the 
remnant, and the ambient density estimates of Toor (1980; $n_{0} = 
0.3$~cm$^{-3}$) and Vink et~al.\ (2000; $n_{0} \sim 0.1$~cm$^{-3}$).  
\cite{wil96} find a somewhat larger ambient density [$n_{0} = 0.52 \pm 0.08 
(d / 1.7\ {\rm kpc})^{1/2}$~cm$^{-3}$], and \cite{dwa98} report a somewhat 
lower density ($n_{0} = 0.05$--0.1~cm$^{-3}$).  Although the remnant is 
interacting with a relatively dense environment in the northwest 
\citep[$n_{0} \approx 1$~cm$^{-3}$;][]{win97}, the average ambient denisty at a 
Galactic height of 430~pc is about 0.06~cm$^{-3}$ \citep{bou90}.  Therefore, 
the average ambient density around \snr\ may be about 0.1~cm$^{-3}$.

The inferred amount of shocked material $M_{s} = 8$--67 
$M_{\odot}$.  However, this range of values is an overestimate because the 
relative abundance of helium used in the best-fit model (Table~\ref{tab2}) 
is the same as the relative abundance of heilium in the solar system.  Since 
\cite{sav82} and \cite{bur00} report much smaller densities of helium in the 
neighborhood of \snr, a more appropriate abundance may be about 0.2 times 
the solar abundance.  In this case, the best-fit amount of shocked 
material $M_{s} = 6.8$~$M_{\odot}$, and the 90\% confidence level range 
is 6--47~$M_{\odot}$.  This range is consistent with the mass estimates of 
Pye et~al.\ (1981; $M_{s} = 5$--15~$M_{\odot}$) and Vink et~al.\ (2000; 
$M_{s} = 8.3 \pm 0.8$~$M_{\odot}$).  However, the results of \cite{dwa98} 
suggest a somewhat smaller mass ($M_{s} = 3$--5) and \cite{wil96} obtain 
a much smaller mass [$M_{s} = 1 (d / 1.7\ {\rm kpc})^{1/2}$~$M_{\odot}$].
With the exeption of the estimate of \cite{wil96}, the mass estimates are 
substantially larger than the Chandrasekhar mass.  Therefore, most of the 
hot shocked material is probably forward-shocked interstellar matter.

Unlike the parameters inferred from the thermal component of the BPL$+$NEI
model, most of the parameters inferred from the thermal component of the
BPL$+$RS model are inconsistent with the results of other observations.  For
example, the inferred velocity of the forward shock $v_{\rm fs} = (16 kT / 3
\mu)^{1/2} = 395$--420~km~s$^{-1}$ and the inferred ambient density $n_{0} =
0.006$--0.014~atoms~cm$^{-3}$.  These estimates are about an order of 
magnitude smaller than the results of previously published analyses.  The 
values inferred for most of the other parameters listed in Table~\ref{tab3} 
are also unrealistic for the Raymond-Smith component of the BPL$+$RS model.  
Therefore, the Raymond-Smith model provides a poor description of the 
thermal emission of \snr.  The differences in the physical properties of the 
Raymond-Smith and nonequilibrium ionization models in XSPEC are that the 
Raymond-Smith model is based on the assumptions that the plasma is in a 
state of collisional ionization equilibrium and that the electron and ion 
temperatures have equilibrated.  Neither of these two conditions is assumed 
for the nonequilibrium ionization model.  In fact, at least one of the 
assumptions is not appropriate for \snr.  \cite{lam96} show that the 
electron temperature is significantly smaller than the temperature of the 
ions in the northwest ($kT_{e} < 0.2 kT_{i}$).  This difference in the 
temperatures could explain why the parameters inferred from the Raymond-Smith 
model are inconsistent with other results.

Table~\ref{tab4} lists estimates of the kinetic energy of the matter in the 
remnant, $E_{\rm kin}$, the thermal energy of the shock-heated protons, 
$E_{kT_{p}}$, and electrons, $E_{kT_{e}}$, the energy in the 
magnetic field, $E_{B}$, and the energy in cosmic rays, $E_{\rm cr}$.  Some 
of these estimates are based on the assumptions that the remnant is in a 
Sedov phase (Hamilton et~al.\ 1983) and that the initial kinetic energy of 
the ejecta 
$E_{0} = 10^{51}$~ergs.  The scaling of the estimates with $E_{0}$ is indicated
in the table.  However, the results for \snr\ may be insensitive to 
the assumption that the remnant is in a Sedov phase.  For example, 
\cite{dwa98} suggest that \snr\ has not yet reached the Sedov phase.  The 
results of their analysis suggest that $M_{s} \approx 4$~$M_{\sun}$, 
$n_{0} \approx 0.07$~cm$^{-3}$, $r_{\rm fs} = 8.5$~pc, and $v_{\rm fs} = 
4000$~km~s$^{-1}$.  If these values (instead of the values in Table~\ref{tab3})
are used to compute the parameters in Table~\ref{tab4}, $E_{\rm kin} = 4 
\times 10^{50}$~ergs and $E_{kT_{p}} = 4 \times 10^{50}$~ergs.  Therefore, 
the same distribution of energy is obtained using the results of \cite{dwa98} 
because the smaller estimates of the amount of shocked material and the 
ambient density are offset by the larger estimates of the shock radius and 
velocity.


\subsection{Nonthermal Emission}
\label{dis-nt}

The high-energy X-ray emission from the northeastern and southwestern rims 
of \snr\ is almost certainly nonthermal.  An attempt to describe this 
emission using only thermal bremsstrahlung emission (Hamilton et~al.\ 1983) was 
unsuccessful \citep{lam98}.  As described in section~\ref{obs}, the 
suggestion that the emission is associated with electrons beamed from an 
unseen central object is untenable.  \cite{rey96} argues convincingly that 
the only plausible explanation of the emission is synchrotron radiation from 
electrons accelerated to energies of about 10--100 TeV.  This conclusion is 
supported by the results of the present analysis, which show that the 
nonthermal spectrum steepens with increasing energy (Fig.~\ref{fig6}).  

Figure~\ref{fig7} displays a  compilation of the radio flux-density results 
of \cite{kun70}, \cite{mil71}, \cite{mil75}, \cite{ste77}, and \cite{rog88}.  
The radio data can be fitted with a power law spectrum: $S_{\nu} = (17.9 \pm 
1.1) (\nu/1\ {\rm GHz})^{-0.57 \pm 0.06}$~Jy.  The results of analyses of data 
obtained using the {\sl IRAS} \citep{are89}, \rosat\ PSPC, \rxte\ PCA, EGRET 
\citep{har99}, CANGAROO \citep{tan98}, and JANZOS \citep{all95}
instruments are also shown.  The figure includes estimates of the photon 
spectra associated with synchrotron radiation, inverse Compton 
scattering of the cosmic microwave background radiation, nonthermal 
bremsstrahlung, and the decay of neutral pions \cite{gai98}.  The 
synchrotron spectrum is extrapolated from the radio data to X-ray energies 
assuming that the electron spectrum is as described in section~\ref{crs}.
The nonthermal bremsstrahlung and $\pi^{0}$ spectra were computed assuming 
the density of protons in the remnant is 0.7~cm$^{-3}$.
The nonthermal X-ray spectrum is consistent with a model of 
synchrotron radiation, but it is not consistent with models of nonthermal 
bremsstrahlung or inverse Compton scattering.  The TeV gamma-ray data can be 
described by inverse Compton scattering.

As illustrated in Figure~\ref{fig7}, the detection of X-ray synchrotron 
emission from \snr\ requires the emission of synchrotron radiation at all 
energies between radio frequencies and X rays.  \cite{win97} report the 
detection of faint diffuse emission ($\approx 1.0$--$2.5 \times 
10^{-17}$~erg~cm$^{-2}$~s$^{-1}$~arcsec$^{-2}$) in the H$\alpha$ band around 
the entire circumference of the remnant.  \cite{rey96} notes that the 
surface brightness of the synchrotron emission is about 
28--32~mag~arcsec$^{-2}$ in the visible band.  This range 
corresponds to a flux of 0.1--$6 \times 
10^{-17}$~erg~cm$^{-2}$~s$^{-1}$~arcsec$^{-2}$ in the H$\alpha$ band.  
Therefore, some of the faint diffuse H$\alpha$ flux may be produced by 
synchrotron radiation from electrons that have energies of about 1~TeV.  The 
measurement of the optical (or infrared) synchrotron flux would be very 
important because it would corroborate the claim that the high-energy X-ray 
emission is synchrotron emission and could help determine the shape of the 
cosmic-ray electron spectrum of \snr.

As shown in Figure~\ref{fig7}, the synchrotron and inverse Compton spectra 
of \snr\ have similar shapes.  Gamma-ray emission is detected near the peak 
of the inverse Compton spectrum.  The electrons that produce this emission 
are the same electrons that produce the emission near the peak of the 
synchrotron spectrum.  Therefore, the gamma-ray and 
nonthermal X-ray emission are produced by electrons that have similar 
energies.  For this reason, the nonthermal X-ray and gamma-ray images are 
expected to be similar.  \citet{tan98} report the detection of gamma rays 
from only the northeastern rim of \snr.  The 95\% confidence level upper 
limit on the gamma-ray flux from the southwestern rim ($1.1 \times 
10^{-12}$~cm$^{-2}$~s$^{-1}$ at energies above 1.7~TeV) is no larger than 
about 40\% of the flux of the northeastern rim [$(4.6 \pm 0.6 \pm 1.4) \times 
10^{-12}$~cm$^{-2}$~s$^{-1}$ at energies above 1.7~TeV].  However, the 
nonthermal X-ray fluxes of the northeastern and southwestern rims are 
comparable.  This discrepancy can be explained if the 
electron spectra, magnetic field strengths, or synchrotron and inverse 
Compton emission-angle distributions of the two rims are different.  Of 
these possibilities, the X-ray data can be used to search for differences in 
the shapes of the electron spectra by searching for differences in the 
shapes of the synchrotron spectra of the two rims.  Figure~\ref{fig8} shows 
a comparison of the nonthermal X-ray spectra of the two rims.  For the 
purposes of this comparison, only the \rosat\ PSPC data were used.  The PSPC 
spectra of each rim were fitted with a model that includes a power law 
component ($dN / dE = F (E / 1\ {\rm keV})^{-\Gamma}$) and the thermal 
component described in Table~\ref{tab2}.  The ratio of the nonthermal X-ray 
fluxes of the southwestern and northeastern rims $f(E) = (F_{\rm SW} / 
F_{\rm NE}) (E/1\ {\rm keV})^{\Gamma_{\rm NE} - \Gamma_{\rm SW}}$.  The 1, 2, 
and 3~$\sigma$ confidence level contours for the parameter space defined by 
$F_{\rm SW} / F_{\rm NE}$ and $\Gamma_{\rm NE} - \Gamma_{\rm SW}$ are 
shown in Figure~\ref{fig8}.  Since the point defined by $F_{\rm SW} / F_{\rm 
NE} = 1$ 
and $\Gamma_{\rm NE} - \Gamma_{\rm SW} = 0$ is well inside the 1~$\sigma$ 
confidence contour, there is no reason to believe that the shapes of the 
nonthermal X-ray spectra of the two rims are different in the energy range 
to which the \rosat\ PSPC is sensitive.  The results of the 
present analysis of the \rxte\ PCA data of the two rims also suggest that 
the shapes of the nonthermal X-ray spectra of the two rims are consistent 
with each other in the energy range to which the PCA is sensitive.  
Figure~\ref{fig8} includes curves along which the ratio $f$ is a constant.  
Here, an energy $E = 0.1$~keV is used to compute the value of $f$, because 
we expect that the synchrotron emission at 0.1~keV and the TeV emission are
produced by electrons that have the same energies.  At 
the 3~$\sigma$ confidence level, the nonthermal flux from the southwestern 
rim at 0.1 keV is greater than 50\% of the nonthermal flux from the 
northeastern rim at this energy.  
At face value, this lower limit seems to be inconsistent with the upper limit 
of 40\% on the ratio of the gamma-ray fluxes.  However, a significant
difference between the electron spectra of the two rims produces a
significant difference in the corresponding inverse Compton spectra, but
only a modest difference in the synchrotron spectra.  Therefore, the modest    
constraint on the ratio of the inverse Compton fluxes corresponds to a poor
contraint on the on the ratio of the synchrotron fluxes of the two rims, and 
the results of the present analysis cannot exclude the possibility that the
differences between the X-ray and gamma-ray images are due to differences
between the electron spectra of the two rims.  Perhaps this
issue will be settled after the TeV and X-ray spectra of the two rims are 
better measured.


\subsection{Cosmic Rays}
\label{crs}

The available radio and X-ray data for \snr\ can be used to infer the 
parameters of the spectrum of the cosmic-ray electrons producing the 
synchrotron emission.  For simplicity, the relativistic electron spectrum is 
assumed to have the form $dn_{e} / dE = A_{e} E^{-\Gamma_{e}} 
\exp(-E/\epsilon_{e})$.  This form is the high-energy limit of Bell's (1978) 
expression: 
%
\begin{eqnarray}
dn/dE 
   & = 
   & A g(E,m,\Gamma,\epsilon)
   \\
   & = 
   & A
      (E + m c^{2})
      (E^{2} + 2mc^{2}E)^{-(\Gamma + 1)/2} 
      \exp({-E/\epsilon}),
\label{equ1}
\end{eqnarray}
%
where $E$ $[= (\gamma - 1) m c^{2}$] is the kinetic energy of the particles 
and the exponential cutoff is added to the formula of Bell.  The 
electron spectral index $\Gamma_{e}$ can be determined from the spectral 
index of the radio data: $\Gamma_{e} = 2\alpha + 1 = 2.14 \pm 0.12$.  
However, estimates of the normalization factor $A_{e}$ and the 
exponential cutoff energy $\epsilon_{e}$ of the electrons cannot be 
uniquely determined from the synchrotron data alone because the 
normalization ($S_\nu \propto A_{e} B^{1 + \alpha}$) and the roll-off 
energy ($E_{\rm roll} \propto \epsilon_{e}^{2} B$) of the synchrotron 
spectrum depend on the strength of the magnetic field.  (Note that $E_{\rm 
roll} = 0.1$~keV in Fig.~\ref{fig7}.) Since the normalization of the 
inverse Compton spectrum produced by the relativistic electrons depends on 
$A_{e}$ (and the known spectrum of the cosmic microwave background 
radiation), but not $B$, the TeV gamma-ray data can be used to determine 
$A_{e}$.  Then the flux and the roll-off 
energy of the synchrotron spectrum can be used to determine $B$ and 
$\epsilon_{e}$, respectively.  For example, if the cosmic-ray electrons 
and the magnetic field of \snr\ have effective volume-filling factors $f_{e}
= f_{B} = 0.25$, if the volume of the remnant $V = \case{4}{3} \pi 
\theta^{3} d^{3} = 5 \times 10^{58}$~cm$^{3}$, and if the inverse Compton 
emission is dominated by scattering of the cosmic microwave background 
radiation, $A_{e} = 8.6 \times 10^{-9}$~cm$^{-3}$~GeV$^{\Gamma-1}$, $B = 
10$~$\mu$G, and $\epsilon_{e} = 20$~TeV.  Similar estimates of the 
magnetic field strength are reported by Tanimori et~al.\ (1998; $B = 6.5 \pm 
2$~$\mu$G), Aharonian \& Atoyan (1999; $B = 10$~$\mu$G), and Dyer et~al.\ 
(2001a, $B = 10$~$\mu$G).  However, these estimates of the magnetic field 
strength seem rather small.  If the field strength at a Galactic height of 
430~pc is similar to the strength in the plane \citep{bou90}, a field of 
10~$\mu$G can be little more than a compressed ambient field.  Yet, the 
large and abrupt change in the radio surface brightness in the regions of 
the bright rims of \snr\ \citep{rey86}, the presence of a radially oriented 
component of the field \citep{rey93}, and estimates of the total energy in 
cosmic-ray particles (described below) suggest that the field has been 
significantly amplified.  Perhaps one or more of the assumptions used to 
determine the strength of the magnetic field is wrong.  In particular, the 
effective volume-filling factor of the magnetic field may be smaller than 
the filling factor of the cosmic rays \citep{jun99}.  In this case, the 
magnetic field strength is underestimated.  For example, if $f_{e} = 0.25$ 
and $f_{B} = 0.1$, $A_{e} = 2.4 \times 
10^{-9}$~cm$^{-3}$~GeV$^{\Gamma-1}$, $B = 40$~$\mu$G, and $\epsilon_{e} 
= 10$~TeV.  This value for the magnetic field lies comfortably between the 
compressed value ($B \approx 10\ \mu$G) and the value obtained using the 
minimum-energy condition ($B \approx 100\ \mu$G).  For the remainder of the 
discussion, it is assumed that $B = 40$~$\mu$G and that the cosmic-ray 
electron spectrum is specified by equation~(\ref{equ1}), where $A_{e}$, 
$\Gamma_{e}$, and $\epsilon_{e}$ are listed in Table~\ref{tab5}.  
The inferred spectrum of cosmic-ray electrons is shown in Figure~\ref{fig9}.

Although there is no evidence that cosmic-ray nuclei are accelerated in \snr\ 
\citep[cf.\ ][]{aha99}, \cite{ell91} suggest that relativistic electrons and 
nuclei are accelerated in a similar manner.  Therefore, we estimate the 
parameters of the cosmic-ray nuclei spectra.  For simplicity, only hydrogen 
and helium nuclei are considered.  The nonthermal spectra of these 
particles are assumed to have the same functional form as equation~(\ref{equ1}).
The spectral indices of the protons and alpha particles are assumed to be 
the same as the spectral index of the electrons ($\Gamma_{p} = 
\Gamma_{\rm He} = \Gamma_{e} = 2.14$).  The exponential cutoff energies 
of the different cosmic-ray particles are assumed to be related by the 
magnetic rigidity of the particles (i.e., $\epsilon_{p} = \case{1}{2} 
\epsilon_{\rm He} = \epsilon_{e} = 10$~TeV).  This relationship is 
appropriate if the maximum energy of the electrons is limited by the escape 
of the particles from the remnant \citep{rey96,dye01}, but it is not 
appropriate if $\epsilon_{e}$ is limited by radiative losses.  The 
values of $A_{p}$ and $A_{\rm He}$ are determined using the value of 
$A_{e}$ and the technique described below.  The ratio of the number 
density of protons to helium nuclei $n_{p}:n_{\rm He} = 1:0.02$ in the 
neighborhood of \snr\ \citep{sav82,bur00}.  If the shocked protons 
and helium atoms are fully ionized, then $n_{e} = 1.04 n_{p}$.  (The 
electron contribution from other nuclei is expected to be small.)  Therefore, 
$n_{p}:n_{\rm He}:n_{e} = 1:0.02:1.04$.  The fractional numbers of 
nonthermal protons and helium nuclei are assumed to be the same as the 
fractional number of nonthermal electrons 
(i.e., $\eta \equiv n_{p}^{\rm cr} / n_{p} = 
n_{\rm He}^{\rm cr} / n_{\rm He} = 
n_{e}^{\rm cr} / n_{e}$).  
Using this relationship and the equations
%
\begin{equation}
n_{p}^{\rm cr} 
   = 
   A_{p} 
   \int_{E_{{\rm min},p}}^{E_{{\rm max},p}} dE g(E, m_{p}, \Gamma_{p}, 
      \epsilon_{p}) 
\end{equation}
%
and 
%
\begin{equation}
n_{e}^{\rm cr} 
   = 
   A_{e}
   \int_{E_{{\rm min},e}}^{E_{{\rm max},e}} dE g(E, m_{e}, \Gamma_{e}, 
   \epsilon_{e})
\end{equation}
%
yields
%
\begin{eqnarray}
A_{p} 
   & = 
   & A_{e} 
      \frac{n_{p}}{n_{e}}
      \frac{\int_{E_{{\rm min},e}}^{E_{{\rm max},e}} dE g(E,m_{e}, 
         \Gamma_{e}, \epsilon_{e})}{
         \int_{E_{{\rm min},p}}^{E_{{\rm max},p}} dE g(E, m_{p}, 
         \Gamma_{p}, \epsilon_{p})} 
   \\
\
   & =
   & 1.1 \times 10^{-6}~{\rm cm}^{-3}~{\rm GeV}^{\Gamma-1},
\end{eqnarray}
%
where 
$E_{{\rm min},e} = 0.58$~keV, 
$E_{{\rm max},e} = \epsilon_{e} = 10$~TeV, 
$E_{{\rm min},p} = \case{3}{16} m_{p} v_{\rm fs}^{2} = 14$~keV, and
$E_{{\rm max},p} = \epsilon_{p} = 10$~TeV.
Similarly, 
%
\begin{eqnarray}
A_{\rm He} 
   & = 
   & A_{e} 
      \frac{n_{\rm He}}{n_{e}}
      \frac{\int_{E_{{\rm min},e}}^{E_{{\rm max},e}} dE g(E,m_{e}, 
         \Gamma_{e}, \epsilon_{e})}{
         \int_{E_{{\rm min},He}}^{E_{{\rm max},He}} dE g(E, m_{\rm He}, 
         \Gamma_{\rm He}, \epsilon_{\rm He})}
   \\
\ 
   & = 
   & 1.0 \times 10^{-7}~{\rm cm}^{-3}~{\rm GeV}^{\Gamma-1}, 
\end{eqnarray}
%
where 
$E_{{\rm min},He} = \case{3}{16} m_{\rm He} v_{\rm fs}^{2} = 57$~keV 
and 
$E_{{\rm max},He} = \epsilon_{\rm He} = 20$~TeV.  
The parameters of the cosmic-ray proton and helium spectra are listed in 
Table~\ref{tab5}, and the spectra are shown in Figure~\ref{fig9}.  These 
parameters and equation~(\ref{equ1}) yield a cosmic-ray injection efficiency 
$\eta = 5 \times 10^{-4}$ if the magnetic field $B = 40~\mu$G.  
This efficiency is assumed to be the same for all particle species.  Our 
assumptions about the shapes of the cosmic-ray spectra and about the 
relative numbers of nonthermal protons and electrons yield a number density
of protons that is 160 times larger than the number density of electrons at
1~GeV (i.e., $dn_{
p}^{\rm cr } / dE = 160\, dn_{e}^{\rm cr} /dE$ at 1~GeV; Fig.~\ref{fig9}).  
Within the rather large uncertainties of our estimates, this ratio is 
consistent with the ratio observed at Earth \citep{mey69}.

The energies in cosmic-ray electrons, protons, and helium nuclei can be 
obtained from equation~(\ref{equ1}) and the particle-spectra parameters 
listed in Table~\ref{tab5} and above: 
%
\begin{eqnarray}
E_{e} 
   & = 
   & V \int_{E_{{\rm min},e}}^{E_{{\rm max},e}} dE E A_{e} 
      g(E,m_{e}, \Gamma_{e}, \epsilon_{e}) \nonumber \\
\
   & =
   & 9 \times 10^{47}\ {\rm ergs,} 
   \\
E_{p} 
   & = 
   & V \int_{E_{{\rm min},p}}^{E_{{\rm max},p}} dE E A_{p}
      g(E,m_{p}, \Gamma_{p}, \epsilon_{p})  \nonumber \\
\  
   & =
   & 1 \times 10^{50}\ {\rm ergs,~and} 
   \\
E_{\rm He} 
   & = 
   & V \int_{E_{{\rm min},He}}^{E_{{\rm max},He}} dE E A_{\rm He}
      g(E,m_{\rm He}, \Gamma_{\rm He}, \epsilon_{\rm He}) \nonumber \\
\
   & =
   & 8 \times 10^{48}\ {\rm ergs.}
\end{eqnarray}
%
The sum of these three energies is $E_{\rm cr} = 1 \times 10^{50}$~ergs.  As 
expected, the cosmic-ray energy is dominated by the energy in cosmic-ray 
protons.  The estimates of the cosmic-ray and magnetic field energies are 
listed in Table~\ref{tab4} for the case in which $B = 40\ \mu$G.  \cite{dye01}
estimate the energy in only cosmic-ray electrons and obtain an energy that
is about an order of magnitude larger than $9 \times 10^{47}$~ergs because 
they used a magnetic field of about $10\ \mu$G instead of the field strength 
of $40\ \mu$G used here.

\cite{dru89} and \cite{mar90} modeled the evolution of a supernova remnant 
using conditions that are quite similar to the conditions appropriate for 
\snr.  These authors assume that 1~$M_{\sun}$ of material is ejected with an 
initial kinetic energy of $E_{0} = 1$--$2 \times 10^{51}$~ergs into an 
ambient medium that has a density $n_{0} = 0.1$~cm$^{-3}$.  The predicted 
results for the case in which the injection efficiency $\epsilon = 10^{-3}$ 
(Fig.~3 of \citealt{dru89} and Fig.~4 of \citealt{mar90}) are consistent 
with our 
estimates of the distribution of energy between thermal particles, cosmic 
rays, and kinetic energy (Table~\ref{tab4}).  Note that the cosmic-ray 
injection efficiency parameter $\epsilon$ of \cite{dru89} and \cite{mar90}, 
which should not be confused with our use of $\epsilon$ for the exponential 
cutoff energy of the cosmic-ray spectra, is computed somewhat differently 
from the injection efficiency parameter $\eta$ used here.  If $B = 40~\mu$G, 
$\eta = 5 \times 10^{-4}$ and the corresponding value of $\epsilon = 1 
\times 10^{-3}$.  Therefore, our results and those of \cite{dru89} and 
\cite{mar90} are consistent if the magnetic field is about 40~$\mu$G.  If 
the magnetic field strength is as small as 10~$\mu$G, $\eta = 2 \times 
10^{-3}$, $\epsilon = 4 \times 10^{-3}$, and the total energy in cosmic rays 
$E_{\rm cr} = 4 \times 10^{50}$~ergs (i.e., the energy is expected to be more 
or less equally distributed between kinetic energy, thermal energy, and 
cosmic rays).  For comparison, Figures~2 and 3 of \cite{dru89} suggest that 
almost all of the energy would be contained in the cosmic rays if $\epsilon 
= 4 \times 10^{-3}$.  Therefore, the results of \cite{dru89} are more 
compatible with a magnetic field of $40~\mu$G than a field of $10\ \mu$G.

The estimates of the parameters of the cosmic-ray spectra of \snr\ have 
important implications for Galactic cosmic-ray acceleration.  It is 
generally accepted that Galactic cosmic rays are accelerated predominantly  
in the shocks of supernova remnants.  If this hypothesis is true, an average 
supernova remnant should transfer a sufficient amount of energy to cosmic 
rays, produce a cosmic-ray proton spectrum that has an appropriate spectral 
index, and accelerate protons to energies high enough to explain the 
properties of the cosmic rays observed at Earth.  Since the average flux of 
cosmic rays in the solar system has been constant within a factor of 2
over the last $10^{9}$~yr \citep{ree83}, the mean rate at which cosmic rays 
are energized should be approximately equal to the rate at which energy is 
lost as cosmic rays diffuse from the Galaxy.  This rate 
\citep[$\dot{E}_{\rm loss} = 0.3$--$1 \times 10^{41}$~ergs~s$^{-1}$;][]{par69,
bla87,dru89,lin92} and the assumption that one supernova remnant 
occurs every 30 yr suggest that an average supernova remnant produces 
0.3--$1 \times 10^{50}$~ergs of cosmic rays over the lifetime of the remnant 
(i.e., 3\%--10\% of the initial kinetic energy of the ejecta is transferred to 
cosmic rays).  The estimate of the cosmic-ray energy in \snr\ at the present 
is already consistent with this range if $B \le 100\ \mu$G.  Therefore, the 
total energy of the cosmic-ray protons in \snr\ may be consistent with the 
energy inferred for the sources of Galactic cosmic rays.

The average spectral index of the relativistic cosmic-ray proton spectra 
produced by the accelerators of Galactic cosmic rays ($\Gamma_{p} = 2.2$) 
is inferred from the spectral index of the cosmic-ray proton spectrum 
observed at Earth \citep[$\Gamma_{p} = 2.80 \pm 0.04$;][]{asa98} and the 
spectral steepening due to the energy-dependent escape of cosmic rays from 
the Galaxy \citep[$\Delta \Gamma = 0.6$;][]{swo90}.  The radio synchrotron 
flux of \snr\ is produced by electrons that have energies of about 1 GeV.   
At these energies the spectral index of the cosmic-ray electrons $\Gamma_{e} 
= 2.14 \pm 0.12$.  Since relativistic protons and electrons are thought 
to be accelerated in a similar manner, the relativistic proton spectrum of 
\snr\ may have a spectral index that is consistent with the index inferred 
for the accelerators of Galactic cosmic rays.

The spectral index of the all-particle cosmic-ray spectrum observed at Earth 
is more or less the same up to an energy of about 3000~TeV.  Above this 
energy, the spectrum gradually steepens to have an index of about 3.
Since the observed cosmic rays that have energies below this ``knee'' 
feature are thought to be Galactic, the Galactic cosmic-ray accelerators are 
expected to be capable of accelerating particles to energies as high as 
3000~TeV.  If the mechanism responsible for the acceleration of Galactic 
cosmic rays depends on the magnetic rigidity of the particles (which is true 
for diffusive shock acceleration in supernova remnants), it may be the case 
that the cosmic-ray particles at 3000~TeV are principally iron and that 
protons are accelerated to energies of only about 100~TeV \citep{lag83}.  
The estimated cutoff energy of the electron spectrum of \snr\ 
($\epsilon_{e} = 10$~TeV), while uncertain, is lower than 100~TeV.  
Since relativistic electrons and protons that have the same energy have the 
same magnetic rigidity, the maximum energy of the protons is expected to be 
the same as the maximum energy of the electrons unless the maximum energy of 
the electrons is regulated by radiative losses.  If the magnetic field $B = 
40$~$\mu$G, an electron with an energy $E = \epsilon_{e} = 10$~TeV 
radiates half of its energy in about 400~yr.  Since this time is less than 
the age of the remnant, the maximum energy of shock-accelerated electrons 
may be limited by synchrotron losses.  In this case, the value of the energy 
$\epsilon_{e}$ represents a lower limit on the exponential cutoff 
energy of the proton spectrum because radiative losses are only important 
for electrons, not nuclei.  Therefore, the maximum energy of the protons may 
be consistent with the expected value of $\epsilon_{p} = 100$~TeV.  
However, if the magnetic field strength $B = 10\ \mu$G, the time required for
a 20~TeV electron to radiate half of its energy (3000~yr) is significantly 
greater than the age of the remnant.  If the magnetic field is this small, the 
maximum energy of the electrons in \snr\ may be regulated by the free escape 
of the particles from the remnant \citep{rey96,dye01}.  In this 
case, the maximum energy of the protons is the same as the maximum energy of 
the electrons and is well below the expected energy of 100~TeV.


\section{Conclusion}
\label{con}

We present the results of a spectral analysis of \rxte\ PCA, \asca\ SIS, and 
\rosat\ PSPC data of the supernova remnant \snr.  These data were fitted with 
several sets of thermal and nonthermal X-ray emission models to 
characterize the global spectral properties of the remnant.  The present 
work represents the first attempt to model both the thermal and nonthermal 
X-ray emission over the entire X-ray energy band from 0.12--17~keV.

The best-fit model includes a nonequilibrium ionization component and a 
broken power law component, which are absorbed through an interstellar 
column density $n_{\rm H} = 5.6 \times 10^{20}$~atoms~cm$^{-2}$.  The 
thermal component is described by an electron temperature $kT_{e} = 
0.6$~keV and an ionization timescale $n_{\rm 0}t = 9 \times 
10^{9}$~cm$^{-3}$~s.  The large relative elemental abundance of silicon 
(10--18 times larger than the relative elemental abundance of silicon in the 
solar system) indicates that most of the X-ray--emitting silicon is 
reverse-shocked ejecta.  The inferred mass of the shocked material is 
several times larger than the Chandrasekhar mass, which implies that most of 
the shocked material is interstellar.  

The nonthermal X-ray emission is well described by a broken power law 
component with a low-energy photon index $\Gamma_{1} = 2.1$ and a high-energy 
photon index $\Gamma_{2} = 3.0$.  Since $\Gamma_{2}$ is significantly larger 
than $\Gamma_{1}$, the broken power law represents an approximation to a 
nonthermal spectrum that is steepening with increasing energy.  This result 
supports previous claims that the nonthermal X-ray emission from \snr\ is 
produced by synchrotron radiation from \vhe\ electrons.  Using both the 
radio and X-ray synchrotron results, the spectrum of relativistic electrons 
in the remnant is inferred to have the form $dn_{e} / dE = A_{e} 
E^{-\Gamma_{e}} \exp(-E/\epsilon_{e})$, with $A_{e} = 2.4 \times 
10^{-9}$~cm$^{-3}$~GeV$^{\Gamma-1}$, $\Gamma_{e} = 2.14 \pm 0.12$, and 
$\epsilon_{e} = 10$~TeV.  The values of $A_{e}$ and $\epsilon_{e}$
are based on the assumption that the mean strength of the magnetic field in 
the synchrotron-emitting region $B = 40\ \mu$G.  This value was arbitrarily 
chosen for the present work because it is between the value obtained if the 
magnetic field is merely a compressed ambient field (10~$\mu$G) and the 
value obtained using the minimum-energy condition (100~$\mu$G).  Other 
published estimates of the field strengths are near 10~$\mu$G, but this 
value seems to be incompatible with the large, abrupt change in the radio 
flux at the bright rims, the detection of a significant radial component of 
the field, and the models of \cite{dru89}.

Since \snr\ is expected to accelerate cosmic-ray nuclei as well as electrons, 
we have also estimated the parameters of the proton and helium spectra.  The 
results suggest that the number density of protons is 160 times larger than
the number density of electrons at
1~GeV.  This ratio is comparable to the ratio observed at Earth.  
Integrating over the nonthermal particle spectra yields an estimate of the 
total energy in cosmic rays $E_{\rm cr} = 10^{50}$~ergs.  This energy and the 
differential spectral index of the electrons at 1~GeV ($\Gamma_{e} = 2.14 
\pm 0.12$) are consistent with the hypothesis that Galactic cosmic rays are 
accelerated predominantly in the shocks of supernova remnants.  However, the 
maximum energy of cosmic-ray electrons in \snr\ is well below 100~TeV 
($\epsilon_{e} = 10$~TeV).  If the maximum energy of cosmic-ray protons 
$\epsilon_{p} = \epsilon_{e}$, the remnant does not accelerate 
particles to the energy of the ``knee'' at about 3000~TeV.  Yet, if $B = 40\ 
\mu$G, the maximum energy of the electrons may be limited by synchrotron 
losses and the maximum energy of nuclei might be significantly larger than 
10~TeV.  Therefore, \snr\ is clearly a significant source of Galactic cosmic 
rays (at least cosmic-ray electrons), but the remnant may or may not be 
capable of accelerating particles to energies as high as the energy of the 
knee.  Similar conclusions are reported for other young Galactic shell-type 
remnants \citep[Allen et~al.\ 1997, 1999;][]{rey99}.

Unlike the X-ray and radio synchrotron emission, TeV gamma-ray emission is 
only observed from the northeastern rim of the remnant.  One possible 
explanation for this discrepancy is that the two rims have different 
electron spectra.  The results of the present analysis are consistent with 
the idea that the electron spectra of the two rims are the same, but the 
possibility that the spectra are different cannot be excluded.  Therefore, 
the nature of this puzzle remains a mystery.


\acknowledgments
We thank Stephen Reynolds for many thoughtful and informative discussions 
about particle acceleration in supernova remnants and the photon emission 
processes of the accelerated particles.  We are grateful for Matthew Baring's 
suggestions about the distribution of energy in a supernova remnant.  We 
thank John Houck for helpful discussions about the thermal emission 
properties of shocked plasmas and for carefully reviewing the manuscript.  
We appreciate the many useful suggestions of the anonymous referee that have 
helped improve the paper.  The present work involved the use of the data 
analysis tools and data archives maintained by the HEASARC at NASA/Goddard 
Space Flight Center.  This work was performed, in part, while G.\ E.\ A.\  
held an NRC-NASA/GSFC Postdoctoral Research Associateship, and G.\ E.\ A.\  
warmly acknowledges Keith Jahoda and the \rxte\ PCA team for their support 
and hospitality during his tenure at NASA/GSGC.  The research efforts of G.\ 
E.\ A.\ are supported in part by the {\sl Chandra} X-Ray Center under contract 
SVI-61010 with the Smithsonian Astrophysical Observatory.

\clearpage


\plotone{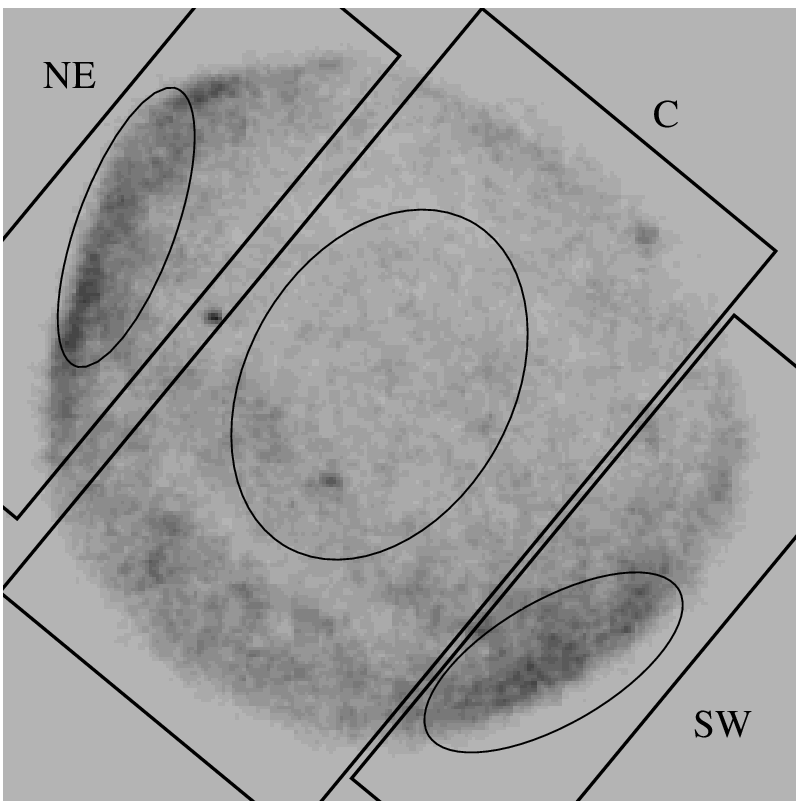}
\figcaption[f1.eps]{A {\sl ROSAT} PSPC image of \snr.  
North is up and east is to the left.  The three rectangles are the 
regions used to extract the PSPC spectra for the northeastern rim (NE), 
center of the remnant (C), and southwestern rim (SW).  The three ovals are 
the regions used to extract the {\sl ASCA} SIS spectra.
\label{fig1}}


\newpage
\vspace*{-0.6in}
\plotone{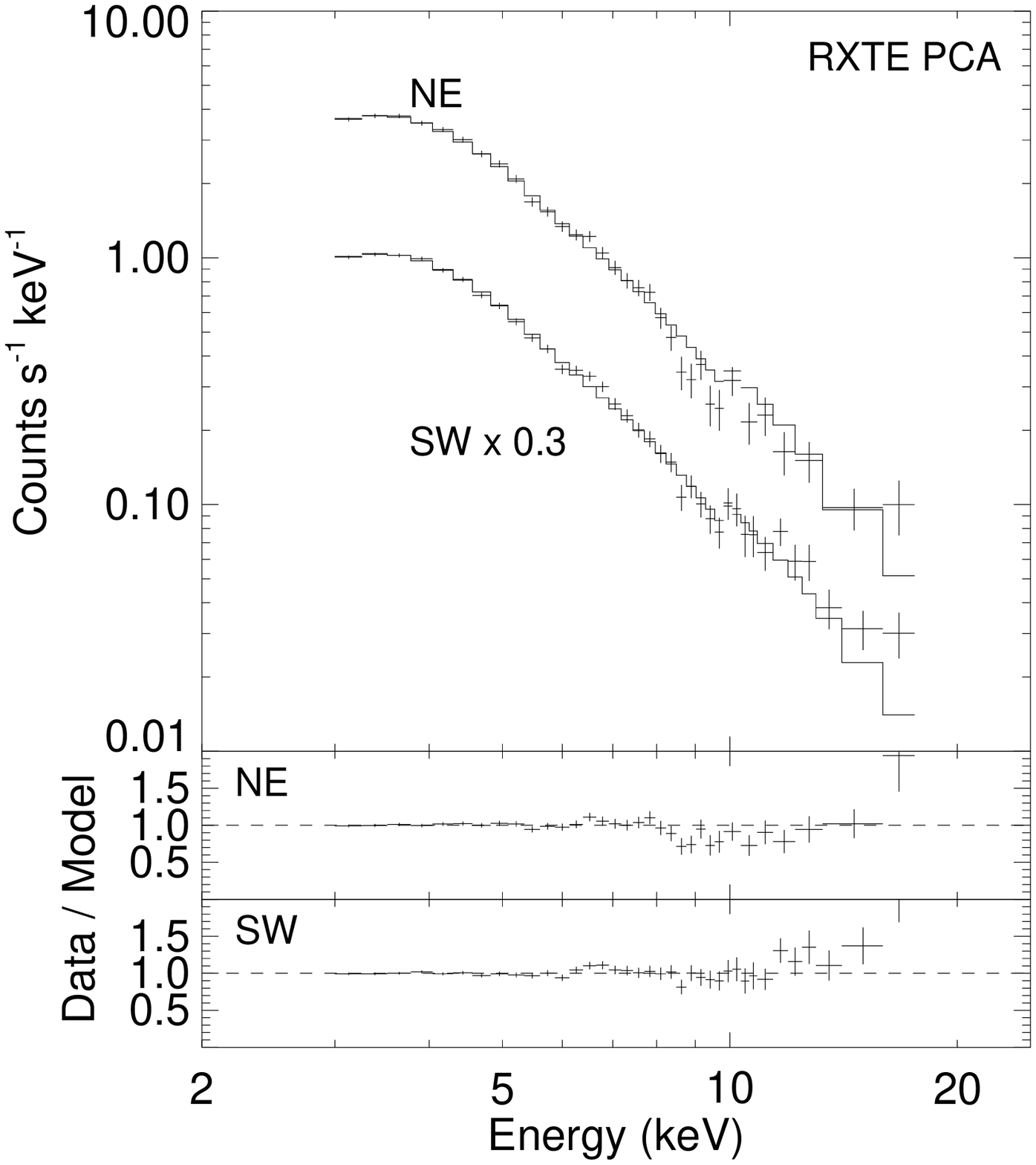}
\figcaption[f2.eps]{\rxte\ PCA data of \snr.  
The top panel includes data for pointings at the the northeastern (NE) and 
southwestern (SW) rims of the remnant.  The entire remnant is in the 
field of view of the PCA at both pointing positions.  The histograms through 
the data show the results of the best-fit model.  The spectral data and model 
of the southwestern pointing have been multiplied by a factor of 0.3 for 
clarity.  The bottom panels show the ratios of the data to the model for both 
pointings.  The model describes the spectra quite well.
\label{fig2}}


\newpage
\vspace*{-0.6in}
\plotone{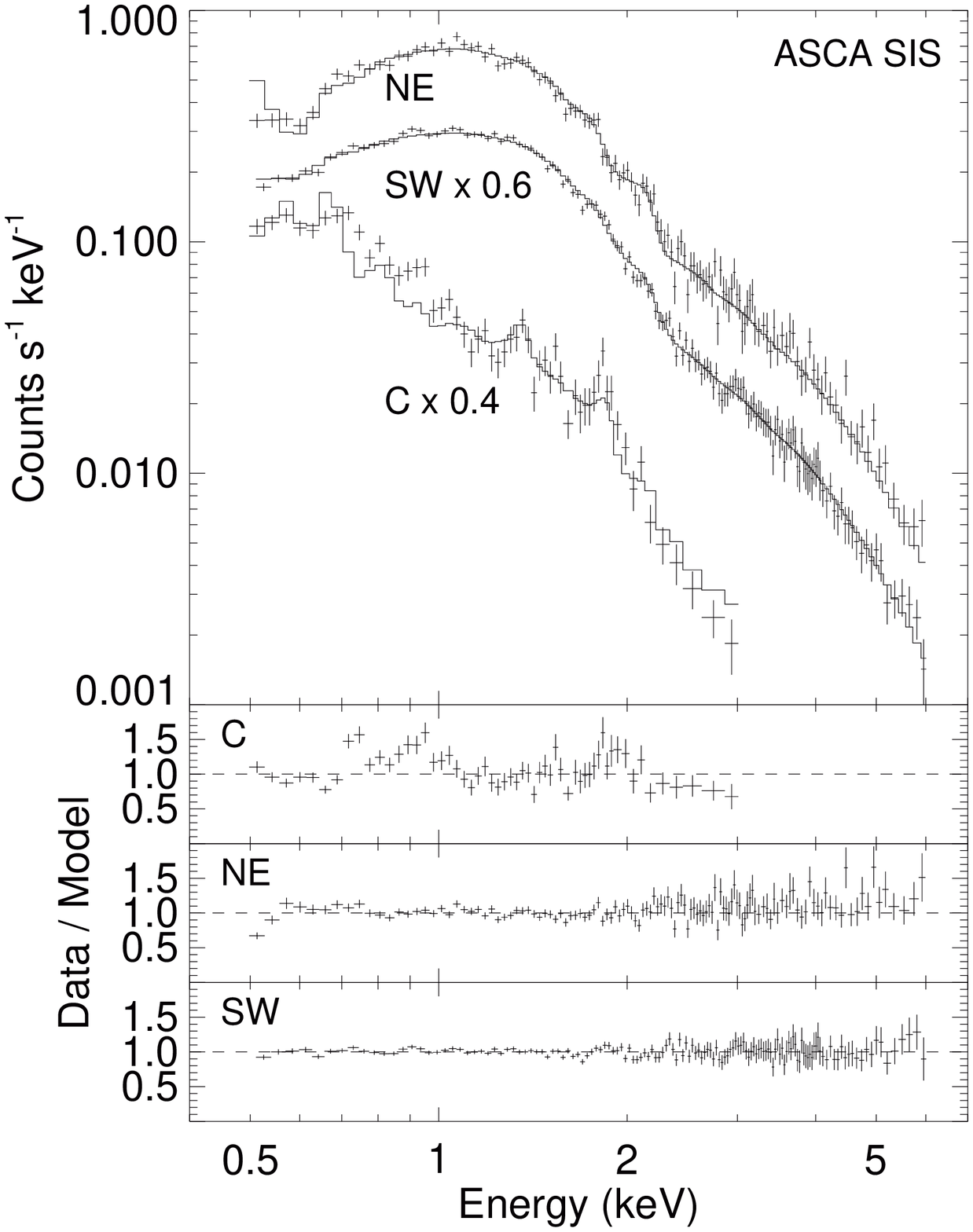}
\figcaption[f3.eps]{\asca\ SIS data of \snr.  The top 
panel includes data for spatially separate regions along the northeastern 
(NE) and southwestern (SW) rims and in the center of the remnant (C).  The 
histograms through the data show the results of the best-fit model.  The 
spectral data and models of the southwestern and central regions have been 
multiplied by factors of 0.6 and 0.4, respectively, for clarity.  The 
bottom panels show the ratios of the data to the model.  In general, the 
model describes the spectra quite well.
\label{fig3}}


\clearpage
\plotone{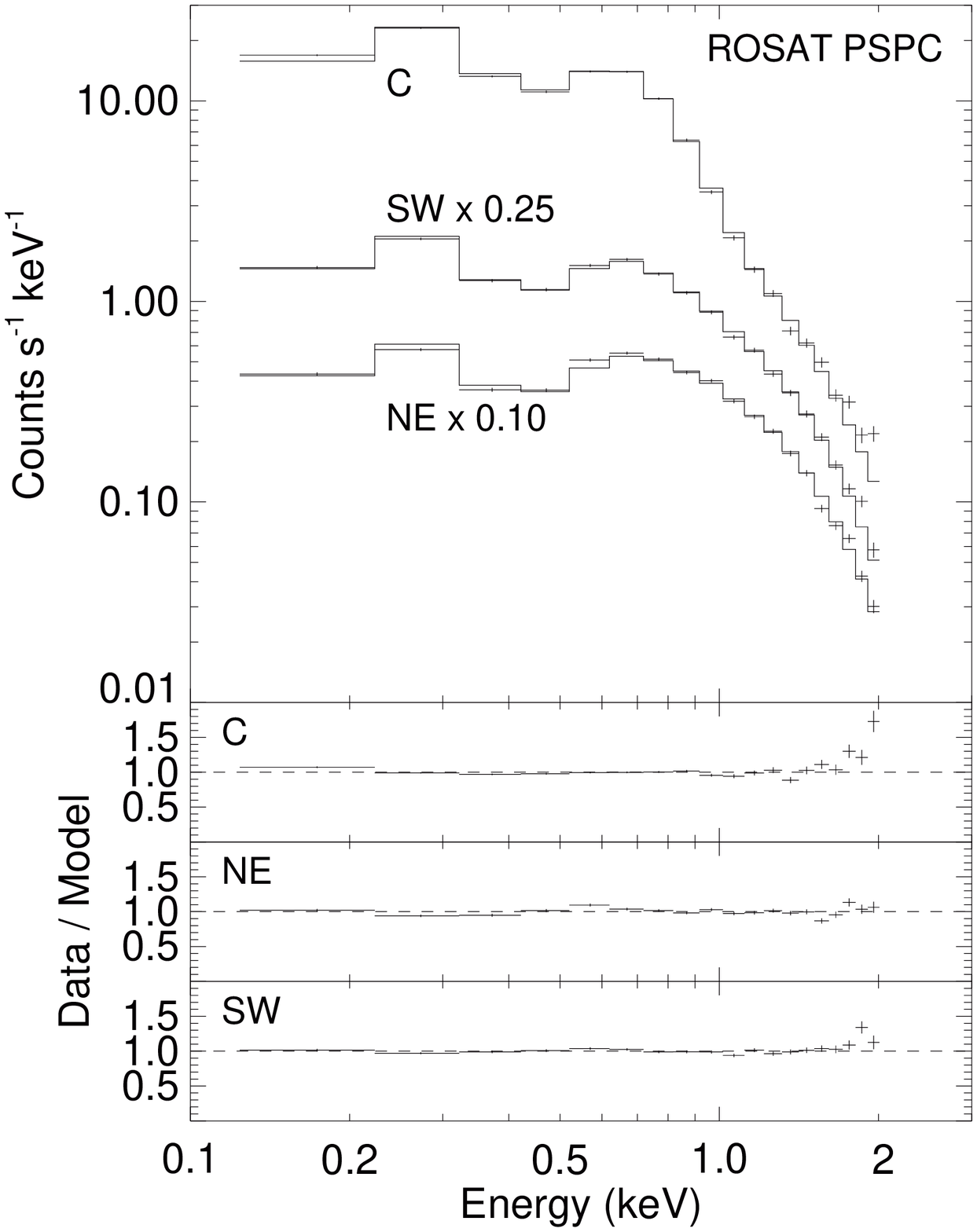}
\vspace*{-0.15cm}
\figcaption[f4.eps]{\rosat\ PSPC data of \snr.  The top 
panel includes data for spatially separate regions in the center of the 
remnant (C) and along the southwestern (SW) and northeastern (NE) rims.  
The histograms through the data show the results of the best-fit model.  The 
spectral data and models of the southwestern and northeastern regions have 
been multiplied by factors of 0.25 and 0.10, respectively, for clarity.  The 
bottom panels show the ratios of the data to the model.  The model describes 
the spectra quite well.
\label{fig4}}


\clearpage
\vspace*{-0.5in}
\plotone{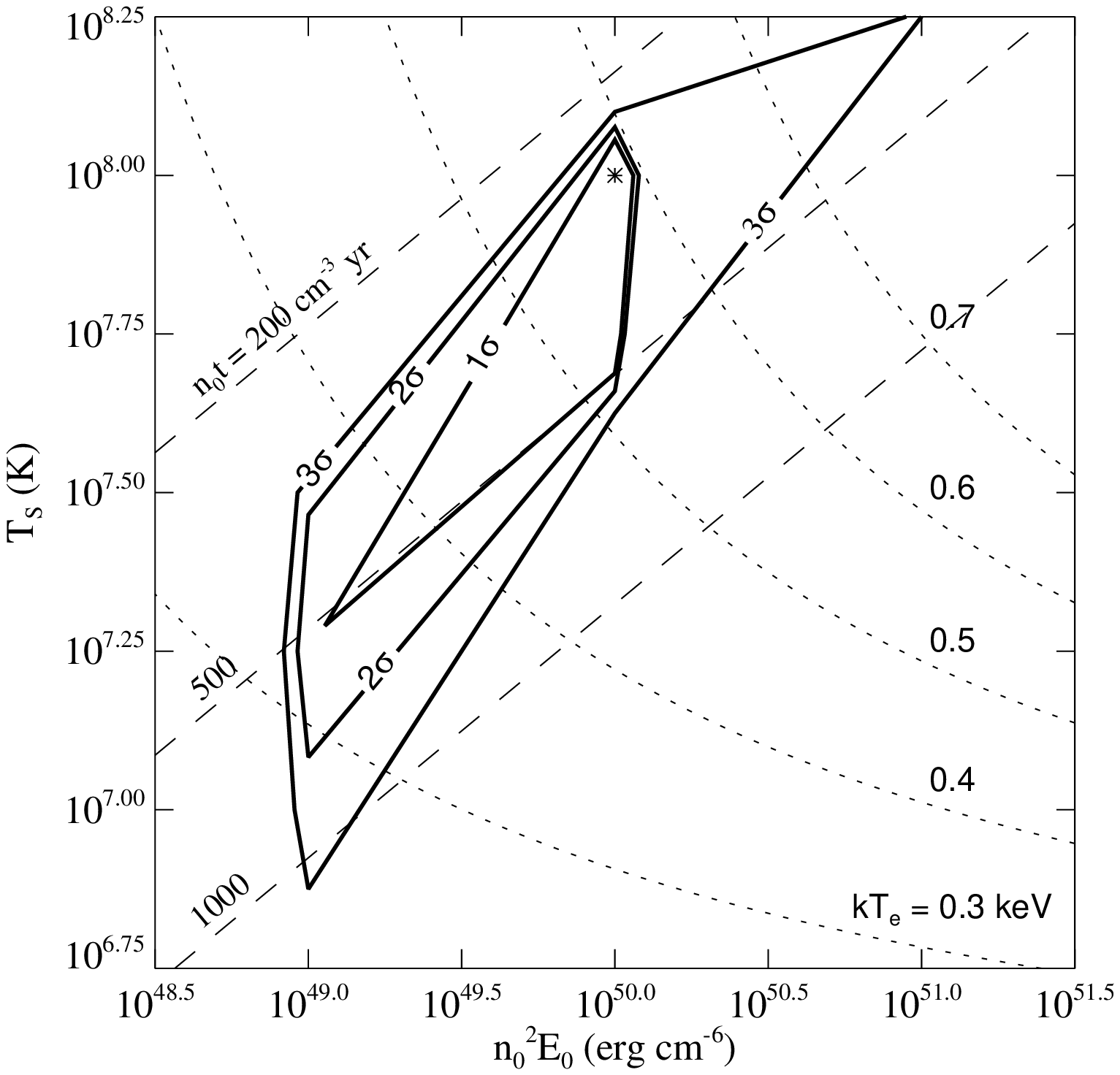}
\vspace*{-0.5cm}
\figcaption[f5.eps]{1, 2, and $3\ \sigma$ confidence level 
contours for the plane of parameter space defined by two of the parameters 
of the nonequilibrium ionization thermal component.  Here $T_{s}$ is 
the temperature associated with the forward shock and $n_{0}^{2}E_{0}$ is 
the product of the square of the ambient density of hydrogen and the initial 
kinetic energy of the ejecta.  The best-fit values of these parameters are 
indicated by the asterisk.  From top to bottom, the three dashed lines 
are lines along which the ionization timescale $n_{0}t = 200$, 500, and 
1000~cm$^{-3}$~yr, respectively.  From bottom to top, the five dotted 
curves are curves along which the characteristic electron temperature 
$kT_{e} = 0.3$, 0.4, 0.5, 0.6, and 0.7~keV, respectively.  
\label{fig5}}


\clearpage
\vspace*{-0.5in}
\plotone{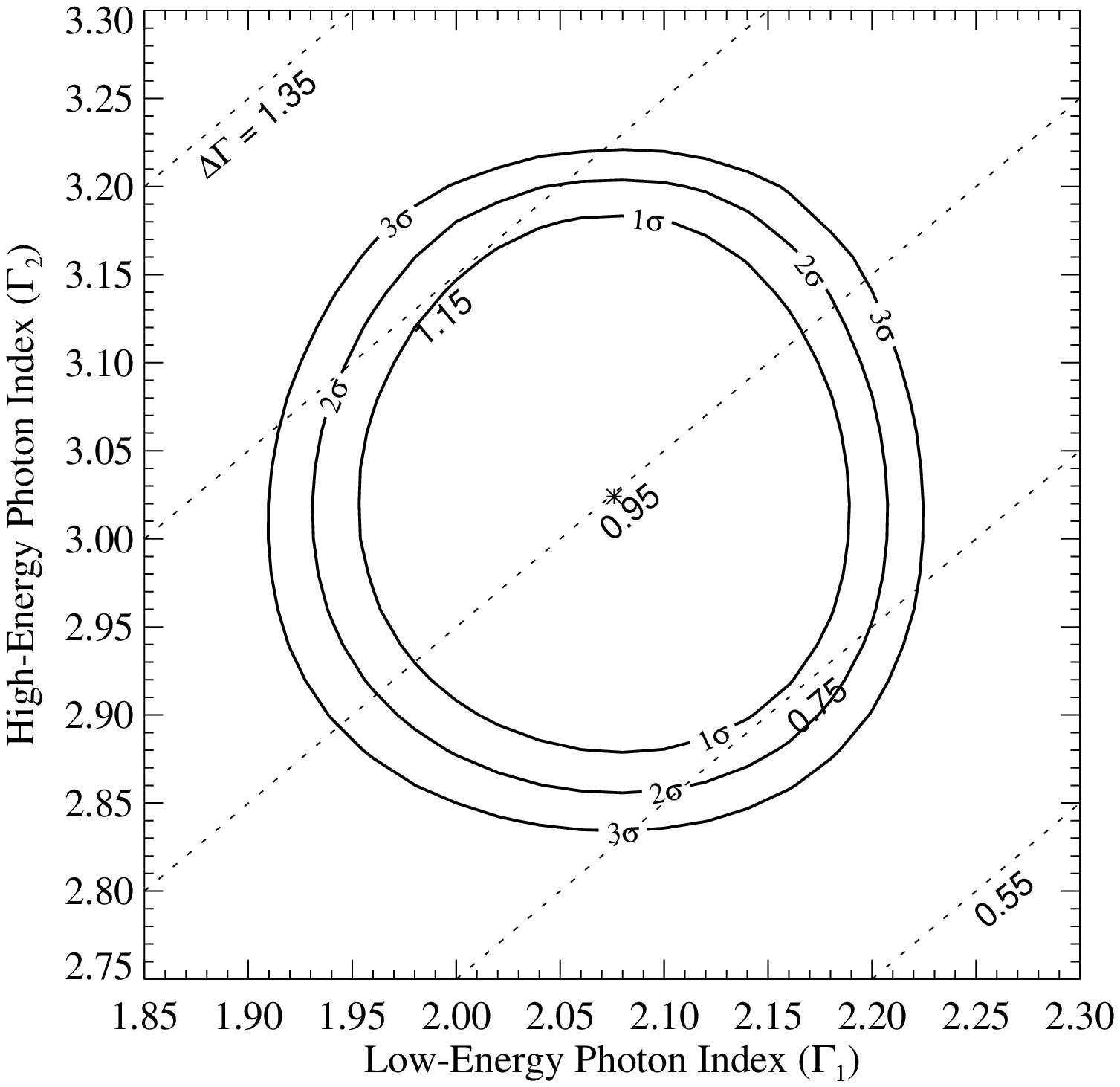}
\vspace*{-0.5cm}
\figcaption[f6.eps]{1, 2, and $3\ \sigma$ confidence level 
contours for the plane of parameter space defined by the photon indices of 
the broken power law.  The best-fit values of the two indices are indicated 
by the asterisk.  The dotted lines are lines along which the difference between 
the two indices $\Delta \Gamma = \Gamma_{2} - \Gamma_{1} = 0.55$, 0.75, 0.95, 
1.15, and 1.35.  The nonthermal X-ray spectrum of \snr\ steepens with 
increasing energy since the line defined by $\Delta \Gamma = 0$ is excluded 
at much more than the $3\ \sigma$ confidence level.  This result supports 
the claim that the nonthermal X-ray emission is synchrotron radiation.
\label{fig6}}


\clearpage
\vspace*{-0.45in}
\plotone{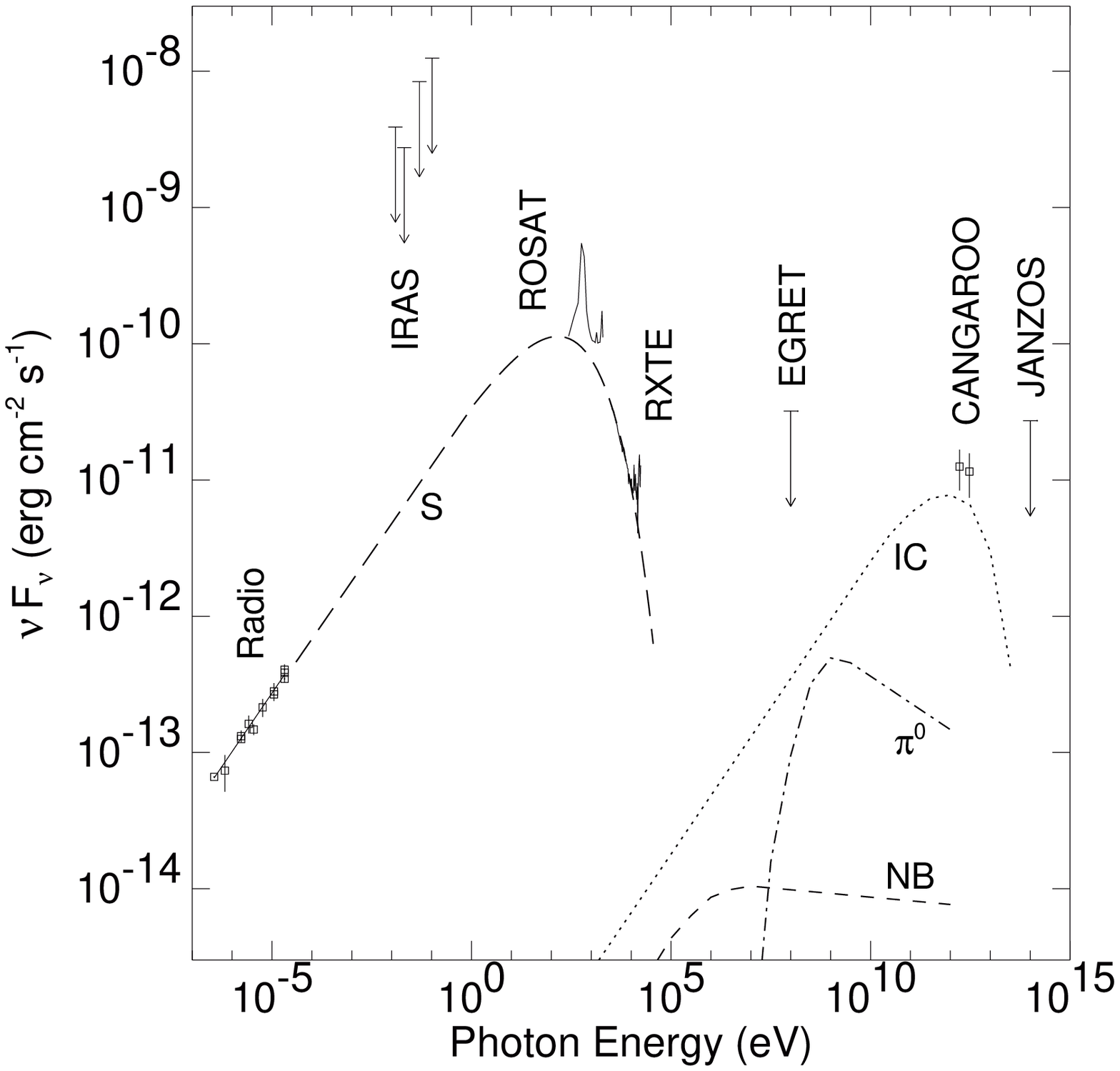}
\vspace*{-0.5cm}
\figcaption[f7.eps]{Radio to gamma-ray photon energy-flux 
spectrum of \snr.  The data, which are labeled vertically, include the radio 
results of Kundu (1970), Milne (1971), Milne \& Dickel (1975), Stephenson, 
Clark, \& Crawford (1977), and Roger et~al.\ (1988), the {\sl IRAS} infrared 
upper limits of Arendt (1989), the \rosat\ PSPC and \rxte\ PCA results of 
this paper, the EGRET gamma-ray upper limit of Hartman et~al.\ (1999, Fig.~3), 
the gamma-ray results of the CANGAROO collaboration (Tanimori et~al.\ 1998), 
and the gamma-ray upper limit of the JANZOS collaboration (Allen et~al.\ 
1995).  The four model spectra are estimates of the photon energy fluxes 
produced by synchrotron radiation (S), inverse Compton scattering on the 
cosmic microwave background radiation (IC), the decay of neutral pions 
($\pi^{0}$), and bremsstrahlung emission of the nonthermal electrons (NB).  
The nonthermal X-ray spectrum is consistent with a model of the synchrotron 
spectrum, but the data are not consistent with the models of nonthermal 
bremsstrahlung and inverse Compton scattering.
\label{fig7}}


\clearpage
\vspace*{-0.5in}
\plotone{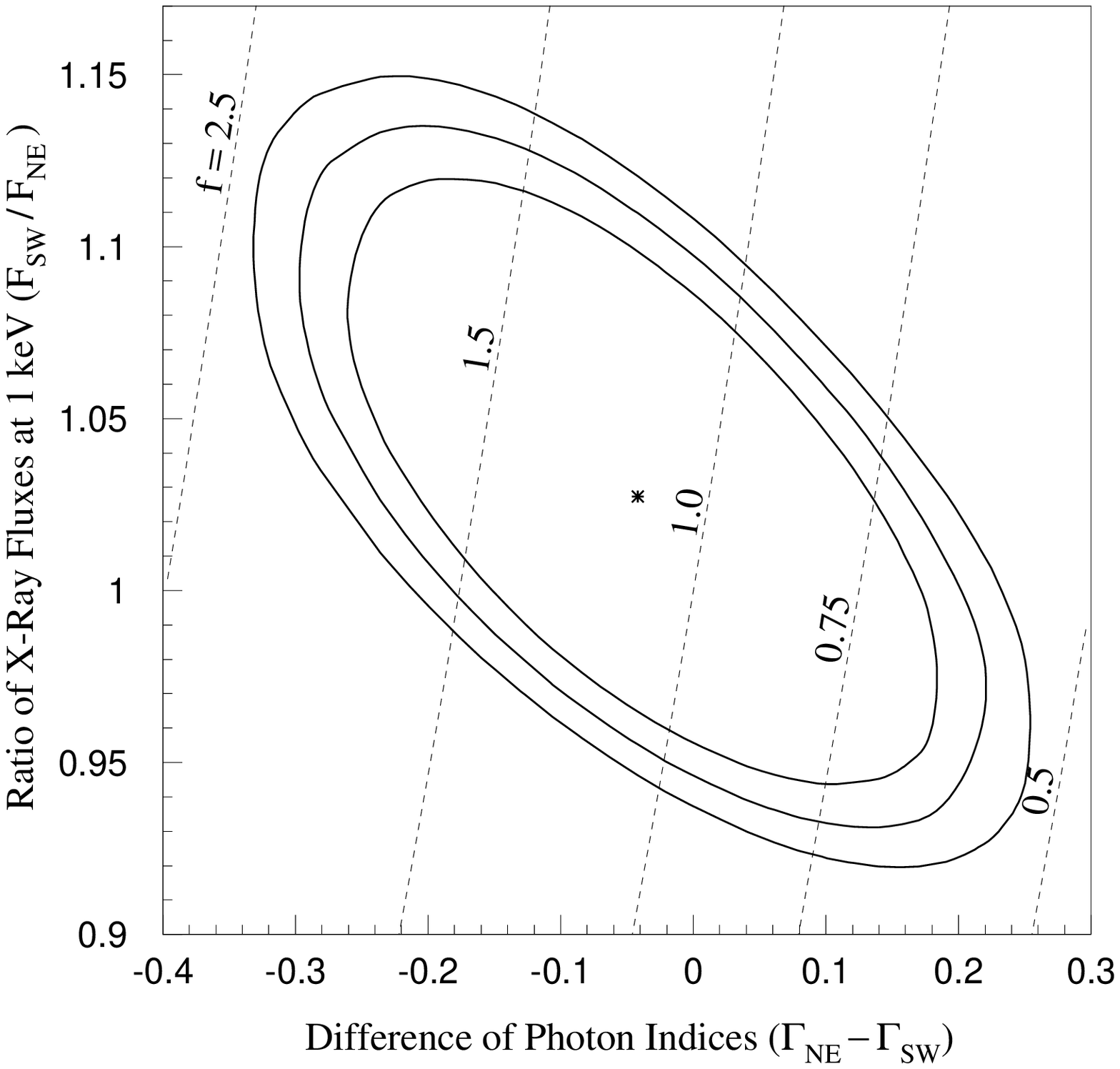}
\vspace*{-0.6cm}
\figcaption[f8.eps]{Comparison of the nonthermal X-ray 
spectra of the northeastern and southwestern rims of \snr\ in the \rosat\ 
PSPC energy band.  Here $\Gamma_{\rm NE}-\Gamma_{\rm SW}$ is the difference 
between the photon indices of the two rims and $F_{\rm SW}/F_{\rm NE}$ is 
the ratio of the nonthermal fluxes at 1~keV.  The best-fit values of the 
two parameters are indicated by the asterisk.  The low-energy nonthermal 
spectra of the two rims are consistent with each other because the point 
defined by $\Gamma_{\rm NE}-\Gamma_{\rm SW} = 0$ and $F_{\rm SW}/F_{\rm NE} 
= 1$ is inside the 1~$\sigma$ confidence level contour.  From the right to 
the left, the five dashed curves are curves along which $F_{\rm SW}/F_{\rm NE} 
= 0.5$, 0.75, 1, 1.5, and 2.5, respectively, at 0.1~keV.  
\label{fig8}}
 

\clearpage
\vspace*{-0.45in}
\plotone{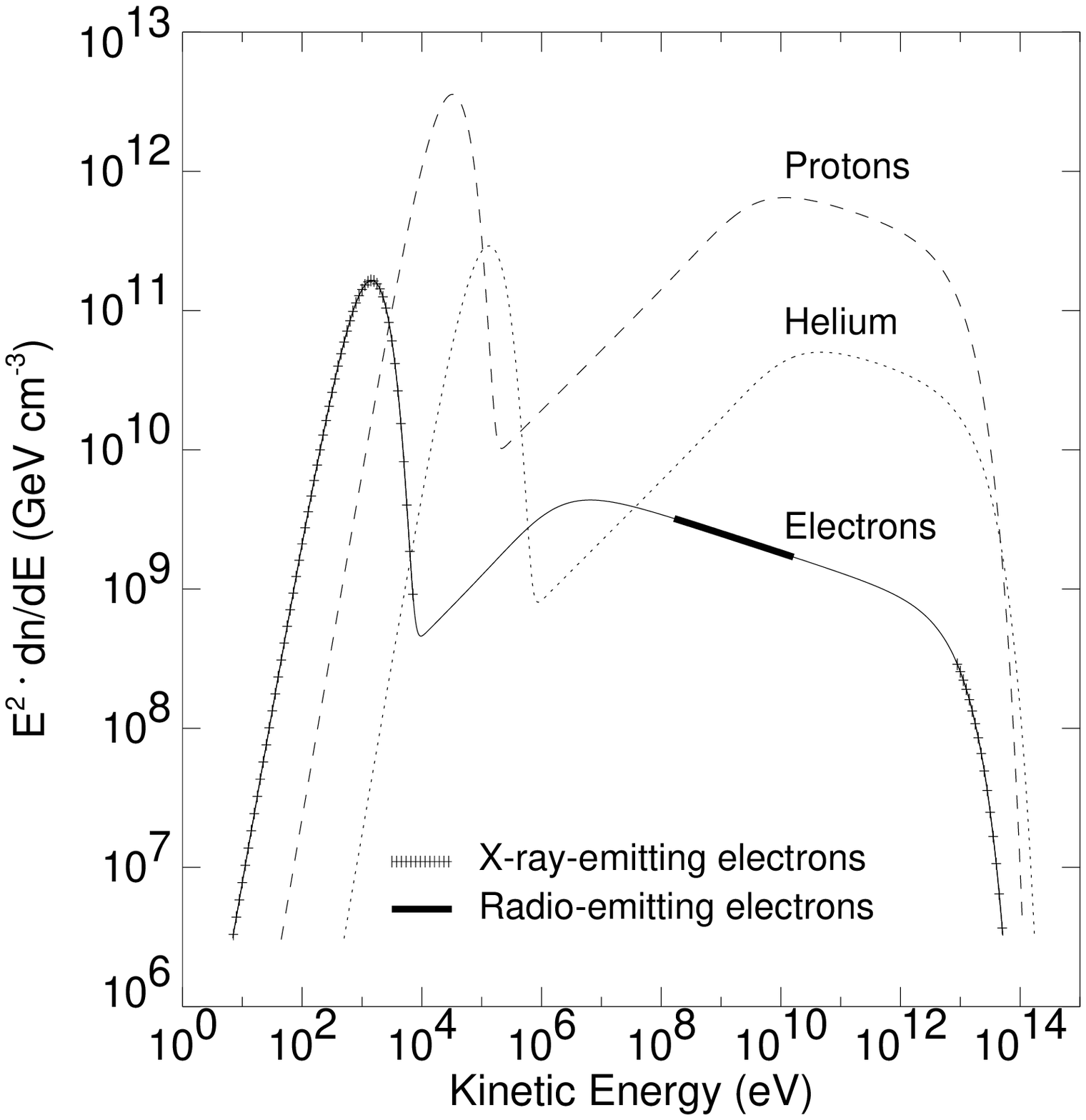}
\vspace*{-0.4cm}
\figcaption[f9.eps]{Estimates of the cosmic-ray electron, 
proton, and helium spectra of \snr.  The low- and high-energy ends of 
the electron spectrum produce the observed thermal and nonthermal X-ray 
emission, respectively.  The GeV electrons produce the observed radio 
emission.  The ratio of the number densities of protons and electrons at 1~GeV 
is about 160, which is consistent with the ratio observed at Earth.  The 
total cosmic-ray energy is dominated by the energy of the cosmic-ray protons.
\label{fig9}}




\clearpage
\begin{deluxetable}{lcccc}
\tablecaption{Results Obtained Using Two-Component Models \label{tab1}}
\tablewidth{0pt}
\tablehead{
   &
   $n_{\rm H}$ &
   Component 1 & 
   Component 2 & \\
\colhead{Model\tablenotemark{a}} &
   ($10^{20}$~atoms~cm$^{-2}$) &
   (Dimensionless or keV) &
   (keV or cm$^{-3}$~s) &
   $\chi^{2}/\nu$
}

\startdata
BPL$+$NEI &
   $5.60^{+0.58}_{-0.51}$ &
   $\Gamma_{1} = 2.08^{+0.11}_{-0.13}$ &
   $kT_{e} = 0.58^{+0.01}_{-0.24}$ &
   1109/729 \\
   &
   &
   $E_{b} = 1.85^{+0.18}_{-0.16}$ &
   $n_{0}t = 8.6^{+7.7}_{-0.7} \times 10^{9}$ &
   \\
   &
   &
   $\Gamma_{2} = 3.02^{+0.16}_{-0.14}$ &
   &
   \\
\cline{1-5}
BPL$+$RS &
   $5.83^{+0.71}_{-0.60}$ &
   $\Gamma_{1} = 2.06^{+0.11}_{-0.12}$ &
   $kT_{e} = 0.162^{+0.009}_{-0.009}$ &
   1202/729 \\
   &
   &
   $E_{b} = 1.90^{+0.15}_{-0.12}$ &
   &
   \\
   &
   &
   $\Gamma_{2} = 3.14^{+0.08}_{-0.09}$ &
   &
   \\
\cline{1-5}
PL$+$NEI &
   $6.48^{+0.52}_{-0.44}$ &
   $\Gamma = 2.51^{+0.04}_{-0.05}$ &
   $kT_{e} = 0.58$\tablenotemark{b} &
   1864/731 \\
   &
   &
   &
   $n_{0}t = 8.6 \times 10^{9}$\tablenotemark{b} &
   \\
\cline{1-5}
Brem$+$NEI &
   $5.13^{+0.65}_{-0.57}$ &
   $kT_{e_{1}} = 2.19^{+0.09}_{-0.10}$ &
   $kT_{e_{2}} = 0.58$\tablenotemark{b} &
   2338/731 \\
   &
   &
   &
   $n_{0}t = 8.6 \times 10^{9}$\tablenotemark{b} &
   \\
\enddata

\tablenotetext{a}{These X-ray spectral models include a power law (PL), a 
broken power law (BPL), a thermal bremsstrahlung model (Brem), a thermal 
model based on the work of Raymond \& Smith (1977; RS), or a
nonequilibrium ionization model (NEI; Hamilton et~al.\ 1983).}

\tablenotetext{b}{The $1\ \sigma$ errors of the characteristic electron 
temperatures, $kT_{e}$, and the ionization time-scales, $n_{0}t$, of the 
NEI model were not determined for these two sets of models, but they are 
expected to be similar to the uncertainties listed for the BPL$+$NEI model.}

\end{deluxetable}


\clearpage
\begin{deluxetable}{lcc}
\tabletypesize{\scriptsize}
\tablecaption{Parameters of the Best-Fit Model\tablenotemark{a} \label{tab2}}
\tablewidth{0pt}
\tablehead{
\colhead{Parameter} &
\colhead{Value} &
\colhead{90\% Interval\tablenotemark{b}}
}

\startdata
\cutinhead{Interstellar absorption component}
$n_{\rm H}$ ($10^{20}$ atoms~cm$^{-2}$) &
   5.60 &
   5.04--6.24
   \\
\cutinhead{Broken power law component\tablenotemark{c}}
$\Gamma_{1}$ &
   2.08 &
   1.94--2.20 \\
$E_{b}$ (keV) &
   1.85 &
   1.66--2.05 \\
$\Gamma_{2}$ &
   3.02 &
   2.86--3.19 \\
Flux(0.1--2 keV) ($10^{-10}$ erg~cm$^{-2}$~s$^{-1}$) &
   1.42 &
   1.31--1.54 \\
\cutinhead{\makebox[5.0in]{Nonequilibrium ionization 
component}}
{$T_{\rm fs}$\ }\ (K) &
   $10^{8}$ &
   $10^{7.25}$--$10^{8}$ \\
{$n_{0}^{2}E_{0}$}\ (ergs~cm$^{-6}$) &
   $10^{50}$ &
   $10^{49}$--$10^{50}$ \\
He\tablenotemark{d} &
   1 &
   \nodata \\
C\tablenotemark{d} &
   1 &
   \nodata \\
N\tablenotemark{d} &
   1 &
   \nodata \\
O &
   1.6 &
   0.8--2.8 \\
Ne &
   0.3 &
   0--0.8 \\
Mg &
   2.0 &
   0--5.9 \\
Si\tablenotemark{e} &
   16 &
   10--18 \\
S\tablenotemark{e} &
   16 &
   10--18 \\
Ca\tablenotemark{e} &
   16 &
   10--18 \\
Fe\tablenotemark{d} &
   1 &
   \nodata \\
Ni\tablenotemark{d} &
   1 &
   \nodata \\
Flux(0.1--2 keV) ($10^{-10}$ erg~cm$^{-2}$~s$^{-1}$) &
   2.53 &
   2.25--2.91 \\
\enddata

\tablenotetext{a}{This model is the ``BPL$+$NEI'' model of Table~\ref{tab1}.}

\tablenotetext{b}{The 90\% confidence level intervals are computed using 
only the statistical uncertainties.}

\tablenotetext{c}{$dF/dE
= K (E/1\ {\rm keV})^{-\Gamma_{1}}$ for $E < E_{b}$ and $dF/dE = K 
(E_{b}/1\ {\rm keV})^{\Gamma_{2}-\Gamma_{1}} (E/1\ 
{\rm keV})^{-\Gamma_{2}}$ for $E \ge E_{b}$.}

\tablenotetext{d}{The relative abundances of the elements helium, carbon, 
nitrogen, iron, and nickel are fixed to be the same as the relative solar 
abundances \citep{and89} of these elements.}

\tablenotetext{e}{The relative abundances of the elements sulphur and 
calcium are fixed to be the same as the relative abundance of silicon.}

\end{deluxetable}


\clearpage
\begin{deluxetable}{lcc}
\tablecaption{Parameters Inferred from the Best-Fit Thermal 
Model \label{tab3}}
\tablewidth{0pt}
\tablehead{
\colhead{Parameter} & \colhead{Value} & 
\colhead{90\% Interval\tablenotemark{a}}
}

\startdata
$kT_{e}$ (keV)                                  & 0.58                   &
   0.31--0.60 \\
$v_{\rm fs}$ (km~s$^{-1}$)                      & 2700                   &
   1000--2900 \\
$r_{\rm fs}$ (pc)                               & 5.9\tablenotemark{b}   &
   5.6--16.3 \\
$d$ ($=r_{\rm fs}/\theta$) (kpc)                
   & 1.4\tablenotemark{b,c} &
   1.3--3.7 \\
$n_{0}t$ ($10^{9}$ cm$^{-3}$~s)                 & 8.6                    &
   7.6--20.6 \\
$t$ (yr)                                        & 870\tablenotemark{b}   &
   760--6300 \\
$n_{0}$ ($=(n_{0}^{2}E_{0}$/$E_{0})^{1/2}$) 
   (cm$^{-3}$) & 0.32\tablenotemark{b}  &
   0.10--0.37 \\
$E_{0}$ ($10^{51}$ ergs)                        & 1.00\tablenotemark{b}  &
   \nodata \\
$M_{s}$ ($M_{\sun}$)                            & 9.7\tablenotemark{b}   &
   8.3--67 \\
\enddata

\tablenotetext{a}{The 90\% confidence level intervals are computed using 
only the statistical uncertainties.}

\tablenotetext{b}{These parameters depend on $E_{0}$, which is assumed to be 
$10^{51}$~ergs.  The parameters $r_{\rm fs}$, $d$, and $t$ scale as 
$(E_{0}/10^{51}~{\rm ergs})^{1/2}$.  The parameter $n_{0}$ scales as 
$(E_{0}/10^{51}~{\rm ergs})^{-1/2}$.  The parameter $M_{s}$ scales as 
$(E_{0}/10^{51}~{\rm ergs})$.}

\tablenotetext{c}{The angular radius $\theta = 15'$.}

\end{deluxetable}


\clearpage
\begin{deluxetable}{lc}
\tablecaption{Inferred Distribution of Energy  \label{tab4}}
\tablewidth{0pt}
\tablehead{
\colhead{} & \colhead{Value} \\
\colhead{Parameter} & \colhead{(ergs)} 
}

\startdata
$E_{\rm kin} = \case{1}{2} M v^{2} \approx \case{9}{32} M_{s} 
   v_{\rm fs}^{2}$ 
   & $4 \times 10^{50}$\tablenotemark{a,b} \\
$E_{kT_{p}} = \case{3}{2} N_{p} kT_{p} \approx \case{3}{8} 
   \pi
   m_{p}
   n_{0}
   r_{\rm fs}^{3}
   v_{\rm fs}^{2}$ 
   & $4 \times 10^{50}$\tablenotemark{b} \\
$E_{kT_{e}} = \case{3}{2} N_{e} kT_{e} \approx 
   2 \pi
   \case{n_{e}}{n_{p}}
   n_{0}
   r_{\rm fs}^{3}
   kT_{e}$
   & $2 \times 10^{49}$\tablenotemark{b} \\
$E_{B}$
   & $3 \times 10^{47}$\tablenotemark{c} \\
$E_{\rm cr}$
   & $1 \times 10^{50}$\tablenotemark{d} \\
\hline 
   \multicolumn{2}{c}{Distribution of cosmic-ray energy 
   $E_{\rm cr}$\tablenotemark{c,d}} \\
\hline
$E_{e}$ 
   & $9 \times 10^{47}$ \\
$E_{p}$
   & $1 \times 10^{50}$ \\
$E_{\rm He}$ 
   & $8 \times 10^{48}$ \\
\enddata

\tablenotetext{a}{The kinetic energy of the unshocked ejecta is neglected.}

\tablenotetext{b}{These parameters scale as $(E_{0}/ 10^{51}~{\rm ergs})$.}
	
\tablenotetext{c}{The magnetic field strength in \snr\ is assumed to be 
40~$\mu$G.}
	
\tablenotetext{d}{Within an accuracy of about 25\%, these parameters scale 
as $(B/40\ \mu{\rm G})^{-1}$.}

\end{deluxetable}


\clearpage
\begin{deluxetable}{lccc}
\tablecaption{Parameters of the Cosmic-Ray Spectra 
   \label{tab5}}
\tablewidth{0pt}
\tablehead{
\colhead{} & \colhead{$A$} & \colhead{} & \colhead{$\epsilon$} \\
\colhead{Particle} & \colhead{(cm$^{-3}$ GeV$^{\Gamma - 1}$)} & 
   \colhead{$\Gamma$} & \colhead{(TeV)}
} 

\startdata
Electrons 
   & $2.4 \times 10^{-9}$
   & $2.14 \pm 0.12$
   & 10
   \\
Protons 
   & $1.1 \times 10^{-6}$
   & 2.14
   & 10
   \\
Helium 
   & $1.0 \times 10^{-7}$
   & 2.14
   & 20
   \\
\enddata

\tablenotetext{\ }{Note.---The cosmic-ray injection efficiency $\eta = 5 \times 
10^{-4}$ and is assumed to be independent of the particle species (see
sec.~\ref{crs}).  The ratio of the number density of cosmic-ray protons to 
the number density of cosmic-ray electrons is 160 at 1 GeV (Fig.~\ref{fig9}).} 
\end{deluxetable}


\begin{thebibliography}{}

\bibitem[Aharonian \& Atoyan(1999)]{aha99}
   Aharonian, F.\ A., \& Atoyan, A.\ M.\ 1999, \aap, 351, 330
   
\bibitem[Allen, Gotthelf, \& Petre(1999)]{all99}
   Allen, G.\ E., Gotthelf, E.\ V., \& Petre, R.\ 1999, in Proc.\ 26th Int.\
   Cosmic Ray Conf.\ (Salt Lake City), 3, 480 (available at
   http://xxx.lanl.gov/abs/astro-ph/9908209) 

\bibitem[Allen et~al.(1997)]{all97}
   Allen, G.\ E., et~al.\ 1997, \apjl, 487, L97 

\bibitem[Allen et~al.(1995)]{all95}
   Allen, W.\ H., et~al.\ 1995, in Proc.\ 24th Int.\ Cosmic Ray Conf.\ (Rome),
   2, 447

\bibitem[Anders \& Grevesse(1989)]{and89}
   Anders, E., \& Grevesse, N.\ 1989, Geochim.\ et Cosmochim.\ Acta, 53,
   197

\bibitem[Arendt(1989)]{are89}
   Arendt, R.\ G.\ 1989, \apjs, 70, 181

\bibitem[Asakimori et~al.(1998)]{asa98}
   Asakimori, K., et~al.\ 1998, \apj, 502, 278

\bibitem[Becker et~al.(1980)]{bec80}
   Becker, R.\ H., Szymkowiak, A.\ E., Boldt, E.\ A., Holt, S.\ S., \&  
   Serlemitsos, P.\ J.\ 1980, \apjl, 240, L33

\bibitem[Bell(1978)]{bel78}
   Bell, A.\ R.\ 1978, \mnras, 182, 443

\bibitem[Blandford \& Eichler(1987)]{bla87}
   Blandford, R.\ \& Eichler, D.\ 1987, \physrep, 154, 1

\bibitem[Bock, Turtle, \& Green(1998)]{boc98}
   Bock, D.\ C.-J., Turtle, A.\ J., \& Green, A.\ J.\ 1998, \aj, 116, 1886

\bibitem[Borkowski et~al.(2000)]{bor00}
   Borkowski, K.\ J., Rho, J., Reynolds, S.\ P., \& Dyer, K.\ K.\ 2000, \apj,
   submitted

\bibitem[Boulares \& Cox(1990)]{bou90}
   Boulares, A., \& Cox, D.\ P.\ 1990, \apj, 365, 544

\bibitem[Burke et~al.(1994)]{bur94}
   Burke, B.\ E., Mountain, R.\ W., Daniels, P.\ J., Cooper, M.\ J., \& Dolat,
   V.\ S.\ 1994, IEEE Trans.\ Nucl.\ Sci., 41, 375

\bibitem[Burleigh et~al.(2000)]{bur00}
   Burleigh, M.\ R., Heber, U., O'Donoghue, D., \& Barstow, M.\ A.\ 2000, 
   \aap, 356, 585

\bibitem[Chevalier \& Raymond(1978)]{che78}
   Chevalier, R.\ A., \& Raymond, J.\ C.\ 1978, \apjl, 225, L27

\bibitem[Drury et~al.(1989)]{dru89}
   Drury, L.\ O'C., Markiewicz, W.\ J., \& V\"{o}lk, H.\ J.\ 1989, \aap,
   225, 179

\bibitem[Dwarkadas \& Chevalier(1998)]{dwa98}
   Dwarkadas, V.\ V., \& Chevalier, R.\ A.\ 1998, \apj, 497, 807

\bibitem[Dyer et~al.(2001a)]{dye01}
   Dyer, K.\ K., Reynolds, S.\ P., Borkowski, K.\ J., Allen, G.\ E., \&  
   Petre, R.\ 2001a, \apj, 551, 439

\bibitem[Dyer et~al.(2001b)]{dye01b}
   Dyer, K.\ K., Reynolds, S.\ P., Borkowski, K.\ J., \& Petre, R.\ 2001b, 
   in AIP Conf.\ Ser.\ 565, Young Supernova Remnants, ed.\ S.\ S.\ Holt \& U.\
   Hwang (New York: Springer)

\bibitem[Ellison \& Reynolds(1991)]{ell91}
   Ellison, D.\ C., \& Reynolds, S.\ P.\ 1991, \apj, 382, 242

\bibitem[Fesen et~al.(1988)]{fes88}
   Fesen, R.\ A., Wu, C.-C., Leventhal, M., \& Hamilton, A.\ J.\ S.\ 1988, 
   \apj, 327, 164

\bibitem[Gaisser, Protheroe, \& Stanev(1998)]{gai98}
   Gaisser, T.\ K., Protheroe, R.\ J., \& Stanev, T.\ 1998, \apj, 492, 219

\bibitem[Galas, Venkatesan, \& Garmire(1982)]{gal82}
   Galas, C.\ M.\ F., Venkatesan, D., \& Garmire, G.\ 1982, \aplett, 22, 103

\bibitem[Hamilton et~al.(1997)]{ham97}
   Hamilton, A.\ J.\ S., Fesen, R.\ A., Wu, C.-C., Crenshaw, D.\ M., \&  
   Sarazin, C.\ L.\ 1997, \apj, 481, 838

\bibitem[Hamilton \& Sarazin(1984)]{ham84}
    Hamilton, A.\ J.\ S., \& Sarazin, C.\ L.\ 1984, \apj, 287, 282

\bibitem[Hamilton, Sarazin, \& Chevalier(1983)]{ham83}
   Hamilton, A.\ J.\ S., Sarazin, C.\ L., \& Chevalier, R.\ A.\ 1983, \apjs,
   51, 115

\bibitem[Hamilton, Sarazin, \& Szymkowiak(1986)]{ham86}
   Hamilton, A.\ J.\ S., Sarazin, C.\ L., \& Szymkowiak, A.\ E.\ 1986, \apj, 
   300, 698

\bibitem[Hartman et~al.(1999)]{har99}
   Hartman, R.\ C.\ 1999, \apjs, 123, 79

\bibitem[Hesser \& van den Bergh(1981)]{hes81}
   Hesser, J.\ E., \& van den Bergh, S.\ 1981, \apj, 251, 549

\bibitem[Jahoda et~al.(1996)]{jah96}
   Jahoda, K., Swank, J.\ H., Giles, A.\ B., Stark, M.\ J., Strohmayer, T.,
   Zhang, W., \& Morgan, E.\ H.\ 1996, \procspie, 2808, 59

\bibitem[Jun \& Jones(1999)]{jun99}
   Jun, B.-I., \& Jones, T.\ W.\ 1999, \apj, 511, 774

\bibitem[Kirshner, Winkler, \& Chevalier(1987)]{kir87}
   Kirshner, R.\ P., Winkler, P.\ F., \& Chevalier, R.\ A.\ 1987, \apjl, 
   315, L135

\bibitem[Koyama et~al.(1997)]{koy97}
   Koyama, K., Kinugasa, K., Matsuzaki, K., Nishiuchi, M., Sugizaki, M., 
   Torii, K., Yamauchi, S., \& Aschenbach, B.\ 1997, PASJ, 49, L7 

\bibitem[Koyama et~al.(1995)]{koy95}
   Koyama, K., Petre, R., Gotthelf, E.\ V., Hwang, U., Matsuura, M., Ozaki, 
   M., \& Holt, S.\ S.\ 1995, \nat, 378, 255

\bibitem[Koyama et~al.(1987)]{koy87}
   Koyama, K., Tsunemi, H., Becker, R.\ H., \& Hughes, J.\ P.\ 1987, PASJ, 39,
   437

\bibitem[Kundu(1970)]{kun70}
   Kundu, M.\ R.\ 1970, \apj, 162, 17

\bibitem[Lagage \& Cesarsky(1983)]{lag83}
   Lagage, P.\ O., \& Cesarsky, C.\ J.\ 1983, \aap, 125, 249 

\bibitem[Laming(1998)]{lam98}
   Laming, J.\ M.\ 1998, \apj, 499, 309

\bibitem[Laming et~al.(1996)]{lam96}
   Laming, J.\ M., Raymond, J.\ C., McLaughlin, B.\ M., \& Blair, W.\ P.\ 
   1996, \apj, 472, 267

\bibitem[Lampton, Margon, \& Bowyer(1976)]{lam76}
   Lampton, M., Margon, B., \& Bowyer, S.\ 1976, \apj, 208, 177

\bibitem[Leahy, Nousek, \& Hamilton(1991)]{lea91}
   Leahy, D.\ A., Nousek, J., \& Hamilton, A.\ J.\ S.\ 1991, \apj, 374, 218

\bibitem[Lingenfelter(1992)]{lin92}
   Lingenfelter, R.\ E.\ 1992, in The Astronomy \& Astrophysics 
   Encyclopedia, ed.\ S.\ Maran (New York: Van Nostrand \& Reinhold 
   Publishers), 65

\bibitem[Long, Blair, \& van den Bergh(1988)]{lon88}
   Long, K.\ S., Blair, W.\ P., \& van den Bergh, S.\ 1988, \apj, 333, 749

\bibitem[Markiewicz et~al.(1990)]{mar90}
   Markiewicz, W.\ J., Drury, L.\ O'C., \& V\"{o}lk, H.\ J.\ 1990, \aap, 236,
   487

\bibitem[Markwardt \& \"{O}gelman(1995)]{mar95}
   Markwardt, C.\ B., \& \"{O}gelman, H.\ 1995, \nat, 375, 40

\bibitem[Mastichiadis \& de Jager(1996)]{mas96}
   Mastichiadis, A., \& de Jager, O.\ C.\ 1996, \aap, 311, L5

\bibitem[Meyer(1969)]{mey69}
   Meyer, P.\ 1969, \araa, 7, 1

\bibitem[Milne(1971)]{mil71}
   Milne, D.\ K.\ 1971, Australian.\ J.\ Phys., 24, 757

\bibitem[Milne \& Dickel(1975)]{mil75}
   Milne, D.\ K., \& Dickel, J.\ R.\ 1975, Australian.\ J.\ Phys., 28, 209

\bibitem[Moffett, Goss, \& Reynolds(1993)]{mof93}
   Moffett, D.\ A., Goss, W.\ M., \& Reynolds, S.\ P.\ 1993, AJ, 106, 1566

\bibitem[Muraishi et~al.(2000)]{mur00}
   Muraishi, H., et~al.\ 2000, \aap, 354, L57

\bibitem[Nomoto, Thielemann, \& Yokoi(1984)]{nom84}
   Nomoto, K., Thielemann, F.-K., \& Yokoi, K.\ 1984, \apj, 286, 644

\bibitem[Ozaki et~al.(1994)]{oza94}
   Ozaki, M., Koyama, K., Ueno, S., \& Yamauchi, S.\ 1994, PASJ, 46, 367

\bibitem[Parker(1969)]{par69}
   Parker, E.\ N.\ 1969, Space Sci.\ Rev., 9, 651

\bibitem[Pfeffermann et~al.(1987)]{pfe87}
   Pfeffermann, E., et~al.\ 1987, \procspie, 733, 519

\bibitem[Plucinsky et~al.(1993)]{plu93}
   Plucinsky, P.\ P., Snowden, S.\ L., Briel, U.\ G., Hasinger, G., \&  
   Pfeffermann, E.\ 1993, \apj, 418, 519

\bibitem[Pye et~al.(1981)]{pye81}
   Pye, J.\ P., Pounds, K.\ A., Rolf, D.\ P., Seward, F.\ D., Smith, A., \&  
   Willingale, R.\ 1981, \mnras, 194, 569

\bibitem[Raymond, Blair, \& Long(1995)]{ray95}
   Raymond, J.\ C., Blair, W.\ P., \& Long, K.\ S.\ 1995, \apj, 454, L31

\bibitem[Raymond \& Smith(1977)]{ray77}
   Raymond, J.\ C., \& Smith, B.\ W.\ 1977, \apjs, 35, 419

\bibitem[Reedy, Arnold, \& Lal(1983)]{ree83}
   Reedy, R.\ C., Arnold, J.\ R., \& Lal, D.\ 1983, Ann.\ Rev.\ Nucl.\ 
   Part.\ Sci., 33, 505

\bibitem[Reynolds(1996)]{rey96}
   Reynolds, S.\ P.\ 1996, \apjl, 459, L13

\bibitem[Reynolds \& Gilmore(1986)]{rey86}
   Reynolds, S.\ P., \& Gilmore, D.\ M.\ 1986, AJ, 92, 1138

\bibitem[Reynolds \& Gilmore(1993)]{rey93}
   ---------.\ 1993, AJ, 106, 272

\bibitem[Reynolds \& Keohane(1999)]{rey99}
   Reynolds, S.\ P., \& Keohane, J.\ W.\ 1999, ApJ, 525, 368

\bibitem[Roger et~al.(1988)]{rog88}
   Roger, R.\ S., Milne, D.\ K., Kesteven, M.\ J., Wellington, K.\ J., \&  
   Haynes, R.\ F.\ 1988, \apj, 332, 940 

\bibitem[Savedoff \& Van Horn(1982)]{sav82}
   Savedoff, M.\ P., \& Van Horn, H.\ M.\ 1982, \aap, 107, L3

\bibitem[Schaefer(1996)]{sch96}
   Schaefer, B.\ E.\ 1996, \apj, 459, 438

\bibitem[Slane et~al.(1999)]{sla99}
   Slane, P., Gaensler, B.\ M., Dame, T.\ M., Hughes, J.\ P., Plucinsky, P.\ 
   P., \& Green, A.\ 1999, \apj, 525, 357 

\bibitem[Slane \& Lloyd(1995)]{sla95}
   Slane, P., \& Lloyd, N.\ 1995, \apjl, 452, L115

\bibitem[Stephenson, Clark, \& Crawford(1977)]{ste77}
   Stephenson, F.\ R., Clark, D.\ H., \& Crawford, D.\ F.\ 1977, \mnras, 180,
   567

\bibitem[Swordy et~al.(1990)]{swo90}
   Swordy, S.\ P., M\"{u}ller, D., Meyer, P., L'Heureux, J., \& Grunsfeld,
   J.\ M.\ 1990, \apj, 349, 625

\bibitem[Tanaka, Inoue, \& Holt(1994)]{tan94}
   Tanaka, Y., Inoue, H., \& Holt, S.\ S.\ 1994, \pasj, 46, L37

\bibitem[Tanimori et~al.(1998)]{tan98}
   Tanimori, T., et~al.\ 1998, \apjl, 497, L25 

\bibitem[Toor(1980)]{too80}
   Toor, A.\ 1980, \aap, 85, 184

\bibitem[Vartanian, Lum, \& Ku(1985)]{var85}
   Vartanian, M.\ H., Lum, K.\ S.\ K., \& Ku, W.\ H.-M., 1985, \apjl, 288, L5

\bibitem[Vink et~al.(2000)]{vin00}
   Vink, J., Kaastra, J.\ S., Bleeker, J.\ A.\ M., \& Preite-Martinez, A.\ 
   2000, \aap, 354, 931

\bibitem[Vink et~al.(1999)]{vin99}
   Vink, J., Maccarone, M.\ C., Kaastra, J.\ S., Mineo, T., Bleeker, J.\ A.\ 
   M., Preite-Martinez, A., \& Bloemen, H.\ 1999, \aap, 344, 289 

\bibitem[Willingale et~al.(1996)]{wil96}
   Willingale, R., West, R.\ G., Pye, J.\ P., \& Stewart, G.\ C.\ 1996, 
   \mnras, 278, 749

\bibitem[Winkler et~al.(1979)]{win79}
   Winkler, P.\ F., Jr., Hearn, D.\ R., Richardson, J.\ A., \& Behnken, J.\
   M.\ 1979, \apjl, 229, L123

\bibitem[Winkler \& Laird(1976)]{win76}
   Winkler, P.\ F., Jr., \& Laird, F.\ N.\ 1976, \apjl, 204, L111

\bibitem[Winkler \& Long(1997)]{win97}
   Winkler, P.\ F., \& Long, K.\ S.\ 1997, \apj, 491, 829

\bibitem[Wu et~al.(1993)]{wu93}
   Wu, C.-C., Crenshaw, D.\ M., Fesen, R.\ A., Hamilton, A.\ J.\ S., \&  
   Sarazin, C.\ L.\ 1993, \apj, 416, 247

\bibitem[Yoshikoshi et~al.(1997)]{yos97}
   Yoshikoshi, T., et~al.\ 1997, \apjl, 487, 65

\bibitem[Zarnecki \& Bibbo(1979)]{zar79}
   Zarnecki, J.\ C., \& Bibbo, G.\ 1979, \mnras, 186, 51P

\end{thebibliography}
\end {document}